\documentclass[numberedappendix,appendixfloats,iop]{emulateapj}
\usepackage{natbib}
\usepackage[utf8x]{inputenc}
\usepackage{amssymb}
\usepackage{amsmath}
\usepackage{graphicx}
\usepackage{setspace}
\usepackage{color}
\def \Ni {\text{Ni}}
\def \rec {\text{rec}}
\def \reci {\text{rec,1}}
\def \recii {\text{rec,2}}
\def \reciii {\text{rec,3}}

\def \gas {\text{g}}
\def \lum {\text{ls}}
\def \cl {\text{cl}}

\def \out {\text{out}}
\def \bb {\text{BB}}

\def \out {\text{out}}

\def \ad {\text{ad}}
\def \in {\text{in}}

\def \sn {\text{SN}}
\def \ion {\text{ion}}
\def \bol {\text{bol}}
\def \diff {\text{diff}}
\def \L {\text{L}}
\def \ls {\text{ls}}
\def \cs {\text{cs}}
\def \rs {\text{rs}}
\def \d {\text{d}}
\def \r {\text{r}}
\def \abs {\text{abs}}
\def \rsout {\text{rs-out}}
\def \rsin {\text{rs-in}}
\def \T {\text{T}}
\def \ff {\text{ff}}
\def \bf {\text{bf}}
\def \erase #1 {}

\begin{document}

\title{Recombination Effects on supernova Light-Curves}
\author{Tamar Faran \altaffilmark{1}, Tomer Goldfriend\altaffilmark{1,2}, Ehud Nakar\altaffilmark{3} and Re'em Sari\altaffilmark{1}}
\altaffiltext{1}{Racah Institute for Physics, The Hebrew University of Jerusalem, Jerusalem, 91904, Israel}
\altaffiltext{2}{Laboratoire de Physique de l'ENS, \'Ecole Normale Sup\'erieure, PSL Research University, Universit\'e Paris Diderot, Sorbonne Paris-Cit\'e, Sorbonne Universit\'es, CNRS; 24 rue Lhomond, 75005 Paris, France}
\altaffiltext{3}{The Raymond and Berverly Sackler School of Physics and Astronomy,
Tel Aviv University, 69978 Tel Aviv, Israel}

\email{tamar.faran@mail.huji.ac.il}

\begin{abstract}
The light curves of type-II supernovae (SNe) are believed to be highly affected by recombination of hydrogen that takes place in their envelopes. In this work, we analytically investigate the transition from a fully ionized envelope to a partially recombined one and its effects on the SN light curve. The motivation is to establish the underlying processes that dominate the evolution at late times when recombination takes place in the envelope, yet early enough so that $^{56}$Ni decay is a negligible source of energy. We consider the diffusion of photons through the envelope while analyzing the ionization fraction and the coupling between radiation and gas, and find that the main effect of recombination is on the evolution of the observed temperature. Before recombination the temperature decreases relatively fast, while after recombination starts it significantly reduces the rate at which the observed temperature drops with time. This behaviour is the main cause for the observed flattening in the optical bands, where for a typical red supergiant explosion, the recombination wave affects the bolometric luminosity only mildly during most of the photospheric phase. Moreover, the plateau phase observed in some type-II SNe is not a generic result of recombination, and it also depends on the density structure of the progenitor. This is one possible explanation to the different light curve decay rates observed in type II (P and L) SNe.
\end{abstract}


\section{Introduction}
Type-II supernovae (SNe) are characterized by prominent hydrogen lines in their spectra, and are believed to be the result of the core-collapse of a massive star ($\geq8 M_\sun$), which has retained most of its hydrogen envelope prior to the explosion. The most common subclass of this group, composing $\sim 70 \%$ of all type-II SNe is the type II-P class \citep{Li11}. These SNe exhibit a phase of almost constant magnitude in their optical light curves termed as the `plateau', which starts 1$-$2 days after the explosion and lasts $\sim$ 100 days.

Observations indicate that the color temperature before the onset of the plateau exceeds $10,000$ K \citep{Faran18}, implying that the H is fully ionized, but during the plateau phase the observed temperature drops to $\sim$5000-6000 K, where most of the H is recombined \citep[e.g.,][]{Elmhamdi03,Bose13}. For that reason it is believed that recombination plays an important role in shaping the long lasting optical plateau.

Recombination in type II SNe was studied by many authors both analytically \citep[e.g.,][]{grassberg71,grass_nad76,popov93,rabinak11} and numerically \cite[e.g.,][]{utrobin07,woosley&kasen09,bersten11,Dessart13}. Nevertheless, there are still important open questions with regards to this process. For example, it is not clear whether the plateau is a generic result of recombination, or whether other processes and physical properties of the progenitor play a role in shaping the light curve. It is also unclear what produces the distribution in the light curve decline rates, which historically lead to the sub-classification to II-P and II-L SNe, and what is the reason that the plateau temperature settles on the narrow range of $5000-6000$K.
These questions and others had not been fully answered yet, because none of the existing analytical models includes all physical processes necessary to accurately model the recombination within the outflow. All of the existing models include the sharp drop in opacity due to recombination, and its effect on photon escape. However, none of them considers the feedback of the opacity drop on the ability of the gas to achieve thermal equilibrium, due to the sharp decrease in the ionization fraction. This feedback is crucial in determining the propagation of the recombination wave through the outflowing gas and thus in setting the observed color temperature and bolometric luminosity. The purpose of this paper is to develop an analytic model for recombination in type II SNe, which includes all of the main physical processes. This model should improve our understanding of the recombination process and provide a good description of the evolution of bolometric luminosity and color temperature in type-II SNe.

The paper is organized as follows. In section \ref{sec:theory} we derive the equations that govern the evolution of the recombination wave in the outflow in any given hydrodynamic profile of a homologous ejecta. For that we first define several critical coordinates within the outflow that simplify the description of the radiation in the envelope. In section \ref{sec:LC} we use these equations to derive the light curve evolution for various hydrodynamic profiles. First, we derive a fully analytic light curve for a power-law hydrodynamic profile. Then we obtain semi-analytic results for a numerical hydrodynamic profile of an RSG explosion. In section \ref{sec:comparison} we compare our model to previous analytic and numerical works. Finally in section \ref{sec:summary} we discuss the implications of our findings and summarize our main results.

\section{Theory} \label{sec:theory}
Our goal is to accurately model recombination in a realistic SN ejecta, and to find the resulting evolution of the bolometric light curve and color temperature. For that we first define three critical locations within the outflow, which determine the observed luminosity and color temperature (denoted as the luminosity, color and recombination shells). We derive the equations that the radiation and the gas at these locations satisfy at any given time, and these enable us to derive the expected observed parameters for any given hydrodynamic profile. In our analysis we ignore any effect of $^{56}$Ni radioactive decay, either as an energy source or as a source of non-thermal photoionization. We also neglect the contribution of the energy released by the recombination process itself. This is justified in Section \ref{sec:comparison}. We consider here only gas with roughly primordial H to He ratio (about 75\% H by mass) and with metallicity that is at most a few times solar. Since both the scattering and absorbtion opacity at the temperatures and densities of interest ($T\sim 10^4$ K and $\rho \lesssim 10^{-12}{\rm ~gr~cm^{-3}}$) are dominated by hydrogen in H rich mixtures, our analysis is insensitive to the metallicity as long as it is not very high.

\subsection{The adiabatic profile of the envelope} \label{sec:hydro_profile}
Assuming a power-law hydrodynamic profile for the envelope, it is enough to parameterize it with two power-law indices and two normalization factors in order to derive the temporal evolution of the light curve. The evolution of the density is given by
\begin{equation}
\rho(r,t)=f_{\rho} r^{-k} t^{k-3} ,
\label{eq:rhoprof}
\end{equation}
where $k$ is a positive constant and $f_{\rho}$ is a constant 
which depends on the initial properties of the progenitor. The relation between the powers of $r$ and $t$ results from the homologous expansion, $r=v(m)t$. Together with $m \propto \rho r^3$, this implies that Eq \eqref{eq:rhoprof} corresponds to a velocity profile of the form:
\begin{equation}
v(m) = \sqrt{\frac{2E}{M}} \Big(\frac{m}{M}\Big)^{-\frac{1}{k-3}} ~,
\label{eq:vprof}
\end{equation}
where $E$ and $M$ are the energy of the explosion and the envelope mass, repectively. At a given time we parametrize the internal energy in the regions that cool adiabatically as
\begin{equation}
E_{\ad}(m,t)=f_{\ad} m^{s} r^{-1} ,
\label{eq:eprof}
\end{equation}
where $s$ is a positive constant and $f_{\ad}$ depends on the initial conditions of the progenitor and the explosion.
$E_{\ad}$ depends on $t$ through $r=v(m)t$. The parameter $s$ relates to the initial conditions of the explosion via the profiles of the velocity and density right after the passage of the shock, $v_{i}(m)$ and $\rho_i(m)$, and before the following rarefaction wave that changes the gas velocity further \citep[]{matzner&mckee}.
This is through the initial thermal energy of each shell following the breakout, $\sim mv_i^2$,  and the adiabatic losses of the radiation dominated gas so $E_{\ad}(m,t) \approx m v_i^2 (\rho/\rho_i)^{1/3}=m v_i^2 (m/\rho_i)^{1/3} r^{-1}$. 

In order to relate the light curve to the explosion energy and the progenitor mass and radius we use the approximations in Eq \eqref{eq:vprof} and

\begin{equation} \label{eq:rho}
\rho (m,r) \approx \frac{(k-3) m}{4\pi r^3}
\end{equation}
to obtain:
\begin{equation}
f_\rho \approx \frac{k-3}{4\pi}(2E)^\frac{k-3}{2} M^{\frac{5-k}{2}}
\end{equation}
and 
\begin{equation}
f_\ad \approx \frac{ERM^{-s}}{2} ~,
\end{equation}
where $R$ is the progenitor's radius.
As we mention above, in the external parts of the stellar envelope a power-law profile provides a relatively good approximation to the structure of a stellar envelope. For typical progenitors with radiative envelopes $k\approx 10$ and $s\approx 0.8$, while for progenitors with convective envelopes, such as RSGs, $k\approx 12$ and $s\approx 0.9$  \citep[][ and references therein]{matzner&mckee}. At inner parts of the expanding envelope a power-law approximation is less accurate. Nevertheless, we find that it is still a decent approximation, and that based on numerical hydrodynamic simulations, $k=6$ and $s=0.9$ are representative values for a fair fraction of the inner envelope.

\subsection{Simplified recombination}
\begin{figure}
\includegraphics[width=1\columnwidth]{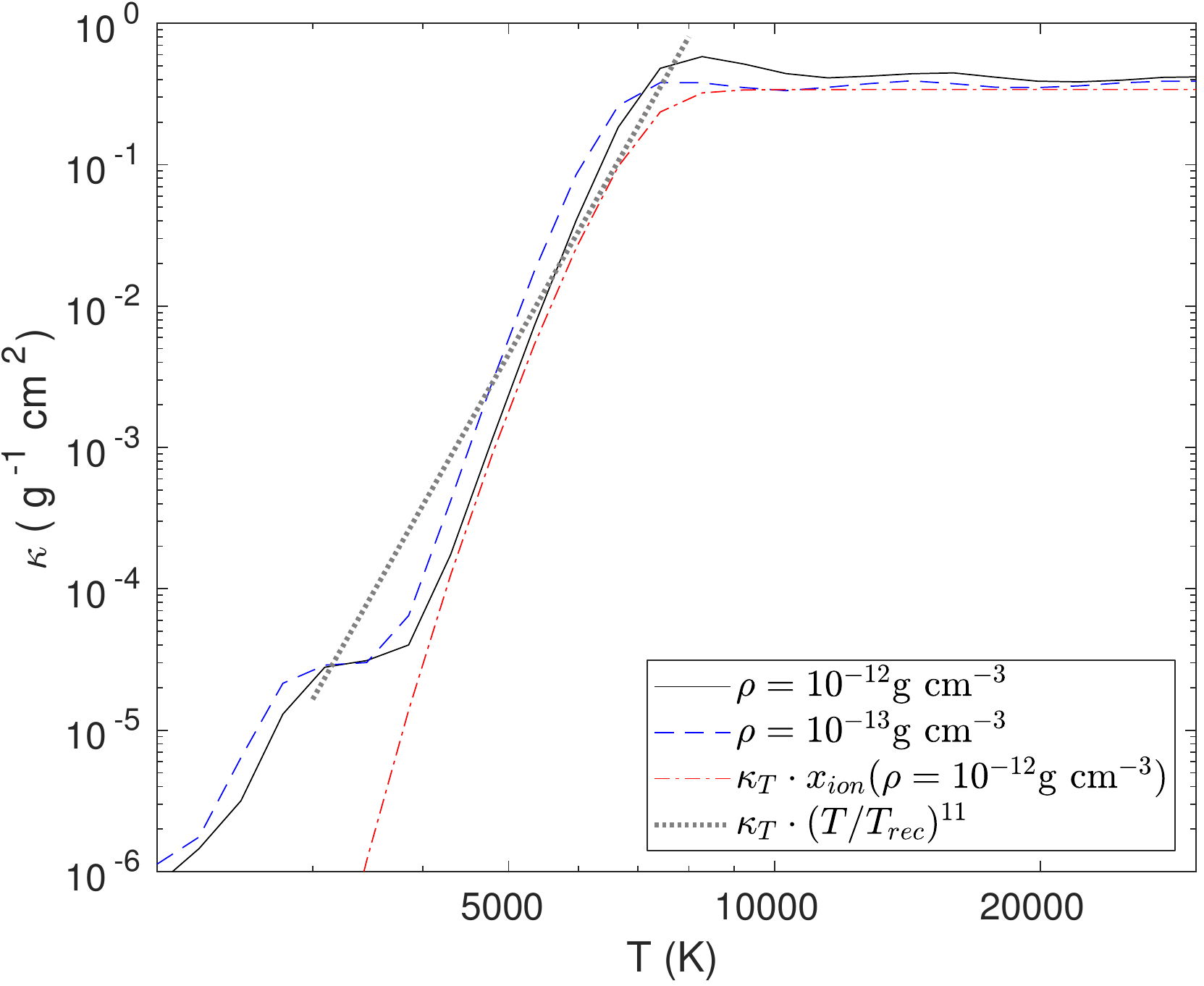}
\caption{Rosseland opacity as a function of the gas temperature, for $\rho = 10^{-12}~\text{g}~\text{cm}^{-3}$ (solid black line) and $\rho = 10^{-13} \text{g}~\text{cm}^{-3}$ (blue dashed line). The approximation $\kappa = \kappa_\T \cdot x_{\ion}$ for $\rho = 10^{-12}~ \text{g}~\text{cm}^{-3}$ using the SAHA equation is plotted as the red dashed-dotted line, and shows a good compatibility with the table opacities. We also plot the power-law approximation which fits well around the temperautre of recombination.} \label{f:rosseland_opacity}
\end{figure}

Since the opacity $\kappa$ in the envelope is dominated by electron scattering, it depends strongly on the ionization fraction of the gas, $x_{\ion}$, which can be determined according to the Saha equation as a function of the gas temperature, $T_\gas$. A good approximation for the opacity is thus given by 
\begin{equation} \label{eq:kappa_xion}
\kappa \approx \kappa_\T \cdot x_{\ion} ~.
\end{equation}
In Figure \ref{f:rosseland_opacity} we plot the Rosseland opacity (using table opacities courtesy of Luc Dessart) as a function of the gas temperature for different densities. The analytic approximation for $\kappa$ using the SAHA equation reproduces well the table opacity values. In order to simplify the expression for $x_\ion$, we expand the Saha equation around $T_{\rec}$ such that $x_{\ion}$ can be written as a broken power law: 
\begin{equation} \label{eq:xion}
   x_{\ion} = \left\{
  \begin{array}{l l}
    1 & \quad   T>T_{\rec}\\
    \left( T/T_{\rec}\right)^{\beta}  & \quad T<T_{\rec}\\
  \end{array} \right. ,
\end{equation}
where in the following we use $\beta=11$. This value of $\beta$ is adequate since for densities of $10^{-12}-10^{-13} ~\text{g}~\text{cm}^{-3}$ the logarithmic derivative of $x_{\ion}$ rises quickly below $T_\rec \approx 7,000-7500$ K up to a value of $11$ around 6000-6500 K, and reaches a maximum of about $15$ at temperatures of 4500-5000 K. We verify that choosing a power-law index in the range $\beta=10-15$ does not significantly affect the results of the analytic solution. We overplot the approximation for $\kappa$ using Eq \eqref{eq:xion} in Figure \ref{f:rosseland_opacity}. The power law expansion fits well around $\sim 0.5$ eV and can therefore reliably reproduce the opacity after recombination starts.

\subsection{Luminosity, color and recombination shells}\label{sec:shells}

During a SN explosion a radiation dominated shock propagates through the envelope of the progenitor star and unbinds it. During its propagation the shock deposits about half of its energy as internal energy in the gas, which later leaks out as radiation during the expansion of the envelope. After the shock crossing and a short phase of acceleration the expanding gas is ionized, radiation dominated and expands homologously. The hydrodynamic profile of the expanding ejecta, where the amount of mass $m$ that is moving at a velocity $v$ drops very sharply with increasing velocity, makes it useful to approximate the outflow as a series of  successive shells. For each shell, there are no considerable changes in the density and velocity over the shell width. The mass of the shell that moves at velocity $v$ is defined as the mass that lies at time $t$ after the explosion between $r=vt$ and the observer and the density within the shell is approximated as a constant $\rho \sim m/r^3$. While the density and the velocity are roughly constant the energy density, gas temperature and the opacity can in general vary significantly across a shell. We define three critical shells within the outflow :

\underline{\it{Luminosity shell}}:  Under our  assumptions the only  source of radiation is  the internal  energy deposited  by the  unbinding SN  shock. The velocity profile and the initially  deposited internal energy implies that the diffusion time to  the observer is dropping significantly with  the shell mass and therefore the luminosity within the outflow can be divided into two regions, inner and outer. In internal regions, which are dense and have a large optical depth, photons  are trapped over a  dynamical time, thereby advected  with the gas. The radiation in these regions is deposited by the SN shock and cools down adiabatically during the  expansion. We therefore can define $E_\ad(m,t)$ as the internal  energy in  shells that are at inner regions. In outer regions photons are free to stream to the  observer over a dynamical  time. The shell that separates the two regions is the {\it luminosity shell} (originally defined in \citealt[]{ns10}, hereafter NS10), which is the location from which radiation streams to the observer over a dynamical time. We denote  this shell with the  subscript `$_{\ls}$'. At radii  larger than the luminosity shell (i.e., in less massive shells) the luminosity  is constant with the radius and  its source is the luminosity shell.  Thus, by finding the luminosity shell we can determine  the observed bolometric luminosity  at any time $t$, which is simply the internal energy in the Luminosity shell divided by $t$. Since the radiation deposited in the luminosity shell by the SN shock is trapped until time $t$ and only then released to the observer, the observed luminosity is simply:
\begin{equation}
 L(t)=E_{\ad,\ls}/t . 
 \label{eq:Luminosity1}
\end{equation}
and is locally related to the energy density, $\varepsilon$ and the optical depth, $\tau$ through
\begin{equation} \label{eq:Luminosity_epsilon}
L(t) = \frac{4\pi r^2 c}{\tau} \frac{d\varepsilon}{dr}
\end{equation}
where $r$ is the distance from the center of the star.

\underline{\it{Color shell}}: Not in all of the shells the radiation and gas are in thermal equilibrium. In inner regions where there are more massive and slower shells, the gas is dense and the coupling between the radiation and the gas is strong enough to allow for thermalization of the radiation. In these regions the radiation is absorbed and re-emitted by the gas, over the time that a typical photon resides in the shell, at least once. The radiation has a blackbody spectrum with a temperature $T_{\bb}$ and its relation with $\varepsilon$ is:
\begin{equation}
 T_{\bb}=(\varepsilon/a)^{1/4} ,
\label{eq:Tbb}
\end{equation} 
where $a$ is the radiation constant. In outer regions the density is low and thermal equilibrium cannot be achieved. In these regions photons are not absorbed and are not generated at a rate that is high enough to maintain thermal equilibrium. As a result, typical photons cannot change their energy and the spectrum of the continuum remains roughly a blackbody  (absorbtion in lines with high optical depth can still take place), but out of thermal equilibrium, namely $\varepsilon<aT^4$. This condition is often denoted as a diluted blackbody (\citealt{Eastman96} and references therein). The two regions are separated by the {\it color shell} which is the outermost shell where the radiation is in a thermal equilibrium. We denote all the quantities in this shell with the subscript `$_{\rm cs}$'.  In order to identify the color shell we define a thermal coupling coefficient
\begin{equation}  \label{eq:eta_def} 
	\eta  \equiv \frac{n_{\text{BB}}}{  t_\d 
	x_{\text{ion}}^2\dot{n}_{\text{ph,ff}}(T_{\text{BB}})}  \quad  ,  
\end{equation}  where
$n_{\text{BB}}=aT_{\bb}^3/3k_B$ is the density of  photons in thermal equilibrium and  $\dot{n}_{\text{ph,ff}}(T) \approx 3.5 \times  10^{36} \text{
s}^{-1} \text{cm}^{-3} \rho^{2} T^{-1/2}$ is the free-free emission rate of fully ionized hydrogen, per unit volume, of photons with energy $h\nu \approx k_B T$. $x_{\text{ion}}$ is the ionization fraction and the factor $x_{\text{ion}}^2$ corrects the photon production rate to a partially ionized gas. $t_\d$ is the diffusion time, and is dependent on $x_\ion$ through $\kappa$ (Eq \ref{eq:kappa_xion}). The parameter $\eta$ defined this way, is the time required to achieve thermal equilibrium ignoring photon  diffusion $\sim n_{\text{BB}}/x_{\text{ion}}^2\dot{n}_{\text{ph,ff}}(T_{\text{BB}})$, divided by the diffusion time, which is the time that a photon resides in the system. This definition is similar to the definition of NS10 with the addition of the ionization fraction which in this study can be smaller than unity.

In our calculations, we assume that free-free is the dominant process for emission and absorption. This is true in low metalicity stars ($\lesssim 0.1$ solar) but not in high metalicity stars ($\gtrsim$ solar). Nevertheless, at the photon energies of interest, $h\nu \approx k_B T_\rec$, bound-free absorption is well described by Kramers' opacity. Therefore, $\kappa_\bf \propto \kappa_\ff \propto \rho T^{-7/2}$, where $\kappa_\bf$ and $\kappa_\ff$ are the bound-free and free-free opacities, respectiveliy. Therefore, when bound-free processes dominate, one simply needs to multiply $\eta$ by the factor $\kappa_\ff/\kappa_\bf<1$, which means that thermalization is achieved faster. However, when the luminosity shell is in thermal equilibrium, this has very little effect on the observed temperature, as explained in NS10.


Regions with $\eta<1$ are in thermal equilibrium while regions with $\eta>1$ are out of thermal equilibrium. Since the ionization fraction and opacity can vary within a shell so does $\eta$. The color shell is the shell where $\eta=1$ at its outer boundary. Beyond this point typical photons cannot change their energy and therefore the observed color temperature, $T_{\cl}$, is determined by the outer boundary of the color shell\footnote{Note that where $\eta>1$ photons are not absorbed in rate that is high enough to keep the gas in equilibrium with the radiation
and in typical RSG explosion Comptonization cannot do that as well. As a result ahead of the color shell not only photons falls out of thermal equilibrium, but also the gas temperature departs from the radiation color temperature}. This is also the location of the so-called\footnote{This definition of the thermalization depth, $\eta=1$, identical to the requirement that the absorption opacity, $\tau_{\nu,abs}$ and the total opacity $\tau_\nu$ for $h\nu =k_BT$ photons satisfy $\tau_{\nu,\abs}\tau_\nu \approx 1$.} "thermalization depth".

\underline{\it{Recombination shell}}: The last property that divides the ejecta is the ionization fraction, which drops very sharply below the recombination temperature $T_{\rec} \approx 7000$ K ($T_\rec$ depends weakly on the gas density). The drop in $x_{\text{ion}}$ leads to a sharp drop in the gas opacity per unit of mass, which is dominated by scattering and thus $\kappa(T) \approx x_{\text{ion}} \kappa_{_{T}}$ (see Figure \ref{f:rosseland_opacity}). In inner regions the temperature is larger than $T_\rec$, the gas is practically fully ionized and $\kappa(T)=\kappa_{_{T}}$. Starting at some time after the explosion the temperature in outer regions falls below $T_\rec$ and then these regions are only partially ionized. The shell in which $T=T_{\rec}$ separates between these two regions. We call it the \textit{recombination shell}, and denote its properties by the subscript `$_{\rs}$'. Because the change in the opacity at $T_{\rec}$ is sharp enough, the energy density and the optical depth profiles change significantly within this shell. This sharp drop in the ionization fraction is the recombination front. The physics of the recombination shell and its propagation determine the observed temperature and in some cases also the observed luminosity. Finding an analytic description of the physics of the recombination shell is one of the the main goals of this paper.
%

Note that a single shell can play several roles at the same time. In practice, shortly after recombination starts the color shell is also the recombination shell and it is possible that in some SNe at late time a single shell is simultaneously the luminosity, color and recombination shell (see next section). Note also that the luminosity is determined by the luminosity shell while the observed color temperature is determined in the color shell, which for typical RSG explosions is usually in front of, or together with, the luminosity shell. Therefore, anything that takes place ahead of the color shell is irrelevant for the bolometric luminosity and observed color. 


\subsection{Schematic description of the propagation of the three shells} \label{sec:shell_propagation}
We first give a schematic description of the propagation of the three shells and the interplay between them. This description is justified quantitatively later. During the entire evolution, from the breakout and until the end of the plateau, all three shells are propagating inwards in lagrangian sense, i.e., the mass of each shell is growing with time. Right after the shock breakout from a typical RSG the color shell is in front of (i.e, at larger radius and with smaller mass) the luminosity shell (NS10). At that time the propagation of the two shells is dominated by diffusion and the mass of the color shell is growing more slowly than that of the luminosity shell. Namely the separation between the two shells is growing. At this time the recombination shell is external to the color shell and is of no interest. 
When the recombination shell reaches the color shell, the two shells coincide and from that point recombination starts affecting the the behaviour of $T_{\cl}$. We define this time as $t_{\rec}$ (later refined into $t_\reci$, $t_\recii$ and $t_\reciii$, see Section \ref{sec:analytic_LC}). As we show in Section \ref{sec:recombination_shell}, from that point onward the recombination shell and the color shell are the same one. Since the luminosity shell is deeper than the color shell at $t_\rec$ only the color temperature evolution is affected at this time while the luminosity evolution remains unchanged. However, the propagation of the color shell at that time is determined by recombination instead of diffusion. As a result it starts propagating inwards faster than the luminosity shell and after some time the recombination and color shells reach the luminosity shell. We define this time as, $t_\L$. As we explain in Section \ref{sec:recombination_shell}, the recombination shell cannot cross the luminosity shell. Instead at $t>t_\L$ a single shell is simultaneously the luminosity, color and recombination shell. At $t_\L$ the propagation of the luminosity shell makes a transition from being dictated by diffusion to being governed by recombination. As a result, this is the time that the observed luminosity starts being affected by recombination. As we show below, depending on the exact progenitor properties, the recombination shell may get to the luminosity shell only after the luminosity shell has crossed most of the SN ejecta, namely $t_\L$ is longer than the plateau duration. In these cases recombination does not significantly affect the plateau luminosity and duration, it only determines the temperature evolution during the plateau. However, in a more accurate treatment of the envelope (as in the semi-analytic calculation in Section \ref{sec:semi_analytic}), one finds that recombination does decrease somewhat the optical depth of shells deeper than the recombination shell, and causes the location of the luminosity shell to recede, thus moderately increasing the bolometric luminosity even before $t_\L$.

\subsection{The structure and evolution of the recombination shell} \label{sec:recombination_shell}

\begin{figure*}
\includegraphics[width=2\columnwidth]{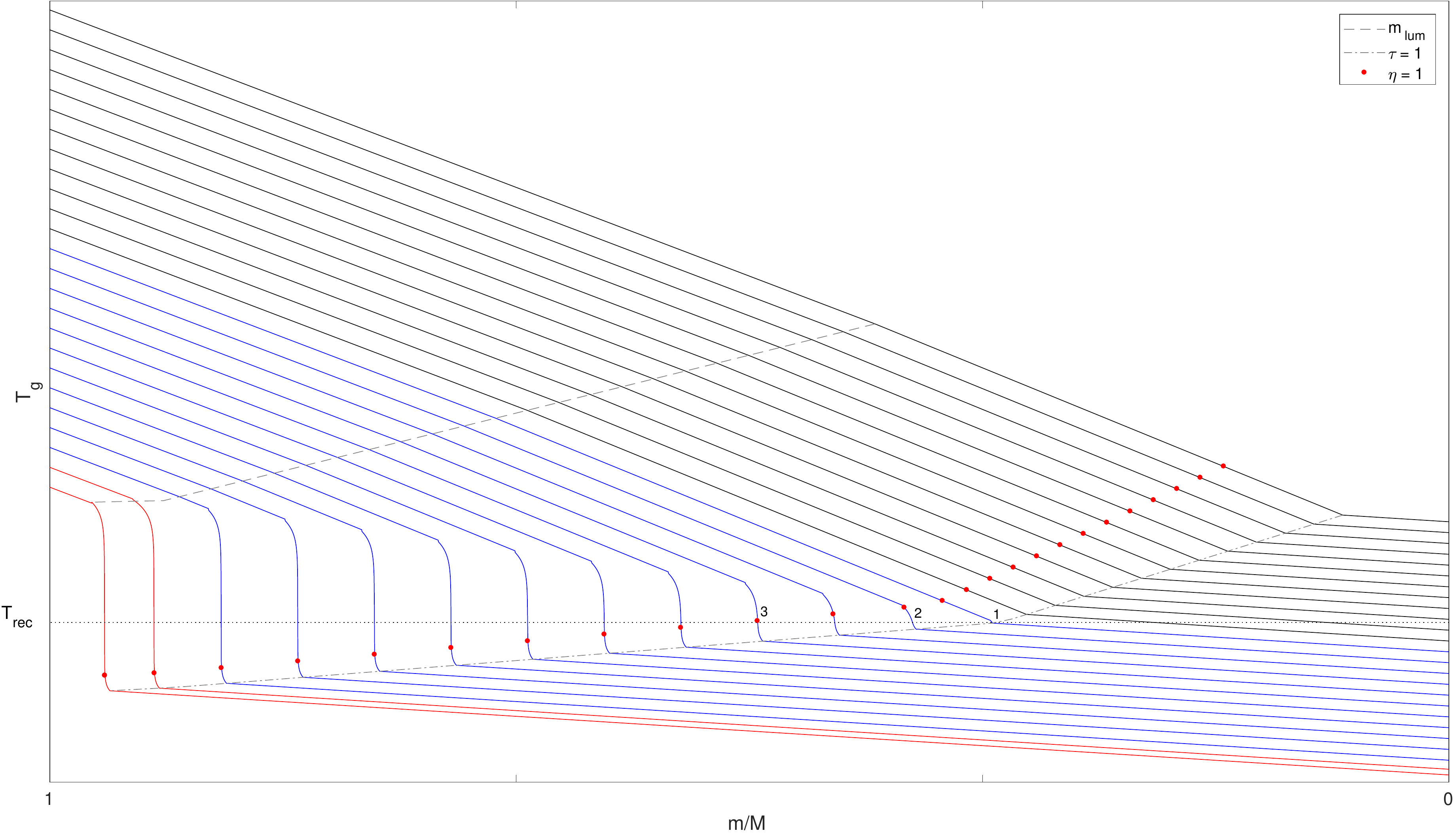}
\caption{A schematic description of the spatial profile of $T_{\gas}$ in the envelope. In black are profiles of $T_\gas$ at $t<t_{\rec,1}$, before the gas at $\tau=1$ started to recombine. The dashed line indicates the location of the luminosity shell, and separates the inner region dominated by adiabatic cooling with that that evolves according to diffusion. The dashed-dotted line follows the location of $\tau=1$, and the red dots indicate the location of $\eta=1$, i.e., where $T_\gas = T_\cl$. Once $T_\gas$ at $\tau=1$ reaches $T_\rec$ (dotted line), the gas starts to recombine and a recombination front is formed which propagates inwards with time (blue curves). As $T_\cl$ reaches $T_\rec$ it starts to evolve much more slowly than before. When the inner boundary of the recombination shell intersects with the luminosity shell, the two shells merge and start propagating together (red curves). The times $t_{\rec,1}$,$t_{\rec,2}$ and $t_{\rec,3}$ are indicated by the markers 1,2 ans 3, respectively.}\label{f:recombination_shell_scheme}
\end{figure*}

Recombination starts in the envelope once $T_\gas$ drops below $T_\rec$. Therefore, the properties of the recombination shell will be dictated by the profile of $T_{\gas}$ throughout the envelope. The power law values we quote in this section are for a single power law profile of $k=12$ and $s=0.9$, for simplicity. Before the onset of recombination, the evolution of $T_{\gas}$ can be divided into three regions witin the outflow. The innermost region lies internal to $m_{\ls}$, in which the change in the radiation energy density profile is governed by adiabatic losses. When the gas is coupled to the radiation, the gas temperature and energy density satisfy
\begin{equation} \label{eq:epsilon_Tg}
T_{\gas} = (\varepsilon /a)^{1/4} ~.
\end{equation}
Together with Eq \eqref{eq:eprof}, we find that in this region $T_{\gas} \propto m^{0.34} t^{-1}$. Beyond the location of the luminosity shell, the energy profile is governed by diffusion, and the gas temperature decreases togther with $\varepsilon$. Using  Eq \eqref{eq:Luminosity_epsilon} and the fact that the luminosity is constant in shells external to the luminosity shell, the dependence of the gas temperature on $m$ and $t$ is $T_{\gas}\propto m^{0.36} t^{-1.04}$, very similar to the profile internal to $m_{\ls}$ for the typical values of $k$ and $s$. $T_{\gas}$ continues to drop with the same power law until $\tau=1$. External to this point, $\varepsilon = L/4\pi cr^2$, and $T_{\gas}$ no longer depends on $\tau$. As we show in the next section, the relation between $\varepsilon$ and $T_{\gas}$ in Eq \eqref{eq:epsilon_Tg} still holds to a decent approximation also beyond $\eta = 1$ (where the gas and radiation decouple). Therefore, at $\tau\leq 1$ the energy density varies as $\varepsilon \propto r^{-2}$ and the gas temperature changes like $T_{\gas}\propto m^{0.05} t^{-0.5}$. The profiles of $T_\gas$ at several epochs before recombination reaches $\tau=1$ are plotted in black in Figure \ref{f:recombination_shell_scheme}. The locations of the luminosity shell and $\tau=1$ are connected by the dashed-dotted line and the dashed line, respectively.

During the adiabatic cooling of the gas, the gas temperature will eventually drop below the recombination temperature of hydrogen (for the gas density in the envelope) and it will start to recombine. As long as recombination takes place only at $\tau<1$, it will have no effect on the profile of $T_{\gas}$ since beyond $\tau=1$, $\varepsilon$ is independent of $\tau$ (which is affected by recombination through the drop in $\kappa$). However, at $\tau>1$, $\varepsilon$ and $T_{\gas}$ will be affected by any change to $\kappa$ caused by the drop in the ionization fraction. We thus refer to the moment the gas at $\tau=1$ starts recombining as the point recombination starts affecting the evolution, and denote it as $t_{\rec,1}$. Before that, recombination has no effect on any of the observed physical parameters.

The profile of $\varepsilon$ inside the recombination shell can again be found using the fact that the luminosity is constant throughout the envelope for $m<m_\ls$, neglecting the recombination energy deposition (this assumption is justified in Section \ref{sec:comparison}). Since the opacity is dominated by scattering, once recombination starts $\kappa = \kappa_\T \cdot x_{\ion}$ and $\tau$ drops rapidly. In order to keep the luminosity constant, $d\varepsilon/dr$ has to change according to Eq \eqref{eq:Luminosity_epsilon}. The outer boundary of the recombination shell is at $\tau=1$, which is where the derivative of the energy density no longer depends on $T_\gas$. The evolution of the recombination shell can thus be found by requiring that $\tau(m_{rs}) = 1$ together with $L(r) = constant$. We find the profile of $T_{\gas}$ inside the recombination shell by treating the regions satisfying $T_{\gas}>T_{\rec}$ and $T_{\gas}<T_{\rec}$ separately. In the first case, the gas is practically fully ionized and in the second case its ionization fraction is set according to Eq \eqref{eq:xion}. 
We define the parameter $\delta$ as the length scale over which the energy density changes inside the recombination shell, namely:
\begin{equation}
\delta = \frac{\varepsilon}{d \varepsilon/dr}~.
\end{equation}
With this definition, the expressions for $L$ and $\tau$ become:
\begin{equation} \label{eq:lum_delta}
L(r) \approx \frac{4\pi r^2 c}{\tau} \frac{\varepsilon}{\delta} \quad , \quad \tau = \kappa \rho \delta ~.
\end{equation}
Demanding that $L(r)$ is constant inside the recombination shell, we find expressions for $\delta(T_{\gas})$ using Eqs \eqref{eq:xion}, \eqref{eq:epsilon_Tg} and \eqref{eq:lum_delta}:
\begin{equation}
\delta (T_{\gas}) = 
 \left\{
  \begin{array}{l l}
 \Big(\frac{T_{\rec}}{T_{\rs-out}}\Big)^{4-\beta} \Big(\frac{T_{\gas}}{T_{\rec}}\Big)^{4} \times r_{\rs}&  
    \quad, T_{\gas}> T_{\rec}
     \vspace{0.5cm} \\
\Big(\frac{T_{\gas}}{T_{\rs-out}}\Big)^{4-\beta} \times r_{\rs} &  
    \quad, T_{\gas}< T_{\rec}  
  \end{array} \right.
  \label{eq:delta}
\end{equation}
where we denote $T_{\rs-out} \equiv T_\gas(\tau=1)$ . The above equation shows that the minimal value for $\delta(T_{\gas})$ is obtained when $T_{\gas}=T_{\rec}$. The inner boundary of the recombination shell satisfies $\delta(T_{\rs-in}) = r_{\rs}$, where $T_{\gas-in}$ is the gas temperature at the inner boundary. This equality gives a relation between $T_{\gas-in}$ and the value of $T_{\rs-out}$:
\begin{equation} \label{eq:Tgas_rec_in}
T_{\rs - in} = \left(\frac{T_{\rec}}{T_{\rs-out}}\right)^{-1 + \beta /4} \times T_{\rec} ~.
\end{equation}
The profiles of $T_\gas$ at $t_{\rec,1}<t<t_\L$ are shown in blue in Figure \ref{f:recombination_shell_scheme}. Once $T_\gas$ at $\tau=1$ is equal to $T_\rec$, it starts to recombine and a steep profile is formed. Inside the recombination shell, where $T_{\gas}>T_{\rec}$, $\delta$ decreases with $T_{\gas}$ (and therefore outwards with $m$), forming a steep profile until reaching a minimun at $T_{\gas}=T_{rec}$. Beyond this point, $\delta$ increases until it reaches $\delta=r_{\rs}$ at $\tau=1$, which marks the outer boundary of the recombination shell. This sharp profile inside the recombination shell is often called the `recombination front'. As $T_{\gas}$ decreases further, the recombination front recedes into deeper layers of the envelope  (in Lagrangian coordinates). When the inner boundary of the recombination shell reaches the luminosity shell, the three shells become one and propagate together, and there is no longer a region that is governed by diffusion. The profiles of $T_\gas$ for $t_\L<t$ are plotted in red in Figure \ref{f:recombination_shell_scheme}.
We note that when the recombination shell is also the luminosity shell, the luminosity is not constant throughout the shell. It is only constant from the point where $t_{\diff}=t$. Nevertheless, we do not treat this subtlety in our derivation, and a more accurate way to derive $\delta$ is brought in Appendix (\ref{sec:delta_acc}).

It is important to note that the dynamics of the recombination shell are (to a good approximation) independent of the thermalization of the gas and radiation. However, as described schematically in Section \ref{sec:shell_propagation}, the ability of the radiation to maintain thermal equilibrium with the gas is a strong function of $x_{\ion}$. Hence, once the recombination shell reaches the color shell, $T_{\cl}$ is expected to be affected by recombination. In Figure \ref{f:recombination_shell_scheme} the locations of $\eta=1$ are designated by the red dots, which satisfy $T_\gas=T_\cl$. Once $T_\cl$ drops below $T_\rec$, $T_\cl$ starts evolving much more slowly and always remains inside the recombination shell. We derive the exact evolution of the shells in Section \ref{sec:LC}.

\subsection{The location of the shells}\label{sec:general_equations}
Here we derive the equations that are satisfied in various regions of the outflow, applying the power-law hydrodynamic profile outlined in Section \ref{sec:hydro_profile}. This enables us to determine the location of the three shells and derive the evolution of the observed luminosity, $L$, and the observed color temperature, $T_{\cl}$ for any given hydrodynamic profile. We are interested in the evolution after recombination starts, but we first give the conditions that take place in the outflow before recombination (i.e., while $T_\gas(\tau=1)>T_\rec$). We do that for completeness (a detailed discussion of this phase is given in NS10) and in order to find the initial conditions for recombination. 

Before recombination starts, the optical depth, $\tau$, and diffusion time, $t_\d$ are given by:
\begin{equation}
\tau(t<t_{\rec}) \approx \frac{(k-3) \kappa_\T m}{4\pi r^2} 
\quad , \quad
t_\d(t<t_{\rec}) \approx \frac{\tau r}{c}.
 \label{eq:tau_nr}
\end{equation}
The luminosity shell is where $t\approx t_\d$ implying:
\begin{equation}
\tau_\ls(t<t_{\rec})\approx c/v_\ls ~,
\label{eq:old_lum_shell}
\end{equation}
which together with Eq  \eqref{eq:Luminosity1} and the initial conditions of the hydrodynamics (after the shock crossing) determines the observed luminosity. The energy density in the outflow in regions where $\tau \geq 1$ is 
\begin{equation} \label{eq:eps}
   \varepsilon(m,t<t_{\rec})\approx \left\{
  \begin{array}{l l}
   \frac{E_{\ad}}{r^3} & \quad   m>m_{\ls}\\
    \\
   \frac{L \tau}{4\pi cr^2} & \quad   m<m_{\ls}\\
  \end{array} \right. ~,
\end{equation}
As we explained in Section \ref{sec:recombination_shell}, recombination starts to affect $\varepsilon$ when $T_{\gas}(\tau=1)=T_\rec$. However, this does not immediately affect the observed properties. The earliest time where $T_\cl$ will be affected by the presence of recombination in the envelope is when $T_\cl = T_{\rs-in}$.
When $T_{\cl} \leq T_{\rec}$, it can be shown that $\tau=1$ and $\eta=1$ are satisfied in the same shell, namely that $m_{\rs} = m_{\cs}$. The reason is that there cannot be an extended region ahead of the recombination shell and behind the color shell where thermal equilibrium is maintained while $\kappa$ is a very strong function of $T_{\gas}$ and the luminosity is constant. The dependence of the opacity on $T_{\gas}$ implies that the gas temperature must drop sharply within the recombination shell, as outlined in Section \ref{sec:recombination_shell}. In this case, the expressions for $\tau$ and $t_\d$ within the recombination shell are:

\begin{equation}
\tau_{\rs}(\delta) \approx \frac{(k-3) \kappa_\T x_{\ion} m_{\rs} \delta}{4\pi r_{\rs}^3} 
\quad , \quad
t_{\d,\rs}(\delta) \approx \frac{\tau_{\rs} \delta}{c} ~.
 \label{eq:tau_wr}
\end{equation}
where $\tau_{\rs}$ is the optical depth inside the recombination shell, $t_{d,\rs}$ is the diffusion time through the recombination shell and $\delta$ was derived in Section \ref{sec:recombination_shell}. The photon temperature also drops, up to the point that the radiation starts losing its coupling with the gas at $\eta=1$. The radiation temperature is then fixed to the value of $T_\gas$ at $\eta=1$, while the temperature of the gas keeps dropping.
When $\eta=1$ is first satisfied inside the recombination shell, $T_\cl$ will have the value of $T_{\gas-in}$ from Eq \eqref{eq:Tgas_rec_in}, therefore $T_{\cl}>T_{\rec}$ and only later $T_{\cl}$ drops below $T_{\rec}$. Nevertheless, the evolution of $T_{\cl}$ will be affected by the profile of $T_{\gas}$ formed inside the recombination shell as soon as $\eta=1$ is satisfied in the recombination shell, due to the dependence of $t_{\d,\rs}$ on the parameter $\delta$.

One point that needs to be addressed is the fact that $T_{\rs-out}$, which determines the value of $\delta$ according to Eq  \eqref{eq:delta}, cannot be found using Eq \eqref{eq:epsilon_Tg}. The reason is that $\tau=1$ lies external to $\eta=1$, meaning that the radiation is out of thermal equilibrium with the gas, and formally, Eq \eqref{eq:epsilon_Tg} does not hold. Instead, the radiation spectrum is a diluted blackbody, with an energy density corresponding to $\varepsilon = d aT_{cl}^4$, where $d$ is the dilution factor. Since $d$ is not easy to estimate, we approximate $\varepsilon$ in the following way. We assume that the gas is in a steady state with the radiation, such that the heating rate by radiation is equal to the cooling rate. The heating rate is given by the expression:
\begin{equation}
\dot{e_h} = daT_\r^4 \int_{0}^{\infty} {c \sigma_{\nu}(T_g)f_{BB,\nu}(T_\r) d\nu} ~,
\end{equation}
where $\sigma_{\nu}$ is the absorption cross section per frequency, $T_\r$ is the radiation temperature, and $f_{BB,\nu}$ is the normalized energy density satisfying $f_{BB,\nu}= u_{\nu}/\int{u_{\nu} d\nu}$ where $u_{\nu} = 4\pi/c B_{\nu}$ and $B_{\nu}$ is the Planck function. From the relation between absorption and emission coefficients, $S_{\nu}=\j_{\nu}/\alpha_{\nu}$ where $S_{\nu}$ is the source function, the cooling rate is
\begin{equation}
\dot{e_c} = aT_{g}^4 \int_{0}^{\infty} {c \sigma_{\nu}(T_{g})f_{BB,\nu}(T_{g}) d\nu} ~.
\end{equation}
By equating the heating and cooling rates, we find the expression for $T_{g}$:
\begin{equation}\label{eq:the_integral}
T_{g} = T_\r d^{1/4} \Bigg(\frac{\int_{0}^{\infty} { \sigma_{\nu}(T_\gas)f_{BB,\nu}(T_\r) d\nu}}{\int_{0}^{\infty} { \sigma_{\nu}(T_\gas)f_{BB,\nu}(T_\gas) d\nu}}\Bigg)^{1/4} ~.
\end{equation}
In the derivation above we assumed that the radiation is a Planckian. This assumption, however, is not accurate. At high frequencies, where $h\nu \geq 13$eV, bound-free processes dominate the absorption and the cross section is much higher than that at $h\nu\sim kT_r$. Therefore, the radiation in those frequencies is in thermal equilibrium with the gas, and the spectral distribution is of a black body with a temperature $T_\gas$. Now, if $T_\r \sim T_\gas$ then the term in the brackets is $\sim 1$. On the other hand, if $T_\gas \ll T_\r$ then the energy at high frequencies is negligible, and the upper limit of the integral is $\sim kT_\r$. We calculate the term in the brackets using opacity tables and find that for $T_\r \geq T_\gas$ it is approximately equal to $1$.
According to Equation (\ref{eq:the_integral}), we can therefore approximate
\begin{equation} \label{eq:e_Tg}
\varepsilon(\eta>1) \approx aT_{\gas}^4 .
\end{equation} 
This approximation elables us to find the dynamics of the recombination shell without dealing with the thermaliztion of the radiation.

Combining the properties of the recombination shell discussed in this section and in Section \ref{sec:recombination_shell}, including Equations (\ref{eq:delta}) and (\ref{eq:e_Tg}), we obtain that for $t \geq t_{\rec}$, the equations governing the dynamics of the three characteristic shells are
\begin{subequations}
\begin{equation}
\left \{
  \begin{array}{l l}
   \tau_\ls = c/ v_\ls & \quad  ,~ t<t_\L ~(\text{i.e., } m_{\ls}>m_{\rs})\\
   m_{\ls}=m_{\rs} & \quad   ,~t>t_\L\\
  \end{array} \right. ,
\label{eq:num_mhat}
\end{equation}
\begin{equation}
\begin{split}
   \eta_\cs= &\left(\frac{T_{\cl}}{7000\text{K}}\right)^{7/2} \left(\frac{\rho_\cs}{10^{-12}gr\cdot \text{cm}^{-3}}\right)^{-3} \times\\
   &\left(\frac{\delta}{3.5\times 10^{12} \text{cm}}\right)^{-2} x_{\ion} ^{-3} = 1~,
\label{eq:num_eta}
\end{split}
\end{equation}
\begin{equation}
L=\frac{E_{\ad}(m_{\ls})}{t} 
=\frac{ (4 \pi)^2 a c T_{\gas}^4  r_{\rs}^5 }{(k-3)\kappa_\T m_{\rs} x_{\ion}(T_{\gas})\delta(T_{\gas}) } ~,
\label{eq:num_lum}
\end{equation}
\begin{equation}
L= 4 \pi a c T_{\rs- \out}^4 r_{\rs}^2 ~,
\label{eq:num_lum_out}
\end{equation}
\begin{equation}
   \tau_{\rs-\out}=1 ~.
\label{eq:num_tau}
\end{equation}
\label{eq:general}
\end{subequations}
The requirement in Eq \eqref{eq:num_eta} states that in the shell where $T_\cl$ is determined (i.e., the color shell), the values of $\delta$ and $\rho$ satisfy $\eta = 1$.
The optical depth, $\tau_\ls$ in Eq \eqref{eq:num_mhat} is computed using $\kappa=\kappa_\T$ as in Eq (\ref{eq:tau_nr}), and $\tau_{\rs-\out}$ in Eq (\ref{eq:num_tau}) is computed using $\kappa = \kappa_\T \cdot x_\ion$ as in Eq \eqref{eq:tau_wr}.
Eq \eqref{eq:num_mhat} states that at $t<t_\L$ the luminosity shell is determined by the requirement $t=t_\d$, while for $t>t_\L$ the recombination shell is also the luminosity shell. Eq \eqref{eq:num_lum} is simply $L \approx 4\pi \epsilon c r^2/\tau$ inside the recombination shell. 
These equations enable us to determine the evolution of the three shells, and thus find $L$ and $T_\cl$ for any given hydrodynamic profile.


\section{Bolometric Luminosity and Observed Temperature} \label{sec:LC}
Here we use the equations derived in the previous section to find the light curve evolution for a typical RSG hydrodynamic profile. First, we use the approximation for the power-law hydrodynamic profile discussed in Section \ref{sec:hydro_profile} with parameters corresponding to a typical RSG progenitor, and apply it to Equation set \eqref{eq:general} to obtain an analytic solution for $L(t)$ and $T_\cl(t)$.
We then present the light curve resulting from applying Equation set \eqref{eq:general} to a numerical hydrodynamical profile  obtained from a numerical explosion of an RSG progenitor, and compare it to the light curve from an analytic broken power-law hydrodynamic profile.

\subsection{Analytic light curve for a power-law hydrodynamic profile} \label{sec:analytic_LC}
For completeness we first provide the evolution of $L$ and $T_\cl$ after the outflow becomes homologous but before recombination starts affecting it. In typical RSGs this phase starts about $5$ hours after explosion and lasts for the first $\sim 10$ days. This phase provides the initial conditions at the time that recombination starts and was discussed in detail in NS10 in terms of $n$, the progenitor polytropic index and $\mu$, a parameter describing the velocity profile. We write the evolution at $t<t_\rec$ in general terms of $k$ and $s$ adding the specific values for $k=12$ and $s=0.9$ (which correspond to $n=1.5$ and $\mu =0.19$, describing the outer regions of the envelope; see NS10 and references therein for details).   

The evolution of the luminosity shell and thus of $L$ before recombination starts is derived from equations  \eqref{eq:vprof}, \eqref{eq:eprof}, \eqref{eq:tau_nr} and \eqref{eq:old_lum_shell}:
$$m_{\ls}(t<t_\rec) \propto t^{\frac{2(k-3)}{k-2}} \approx t^{1.8} ,$$
\begin{equation}
 L(t<t_\rec) \propto t^{-\frac{2(1-s)(k-3)}{k-2}} \approx t^{-0.17} .
 \label{eq:lum_old}
\end{equation}
Equations \eqref{eq:Tbb}, \eqref{eq:eta_def}, \eqref{eq:eps} and the requirement $\eta_\cs =1$ implies that the evolution of the color temperature and the color shell mass are   
$$T_{\cl}(t<t_\rec) \propto t^{\frac{2(k-3)(6ks-11k-4s+14)}{(k-2)(17k-23)}}\approx t^{-0.56} , $$
\begin{equation}
m_{\cs}(t<t_\rec) \propto t^{-\frac{2(k-1-3sks)}{k-2}} \approx t^{1.33} .
\label{eq:dawn_cl}
\end{equation}
The normalizations of all these equations depends on the progenitor and on the explosion properties and are given in NS10. 

To understand the dynamics of $T_\cl$ and $T_\gas$, it is necessary to define three different timescales:
\begin{equation*}
\begin{array}{l l}
t_{\rec,1} :  & \quad \text{when } T_\gas (\tau=1) = T_\rec 
\vspace{0.2cm}\\
t_{\rec,2} : & \quad \text{when } T_\cl = T_{\rs-in}
\vspace{0.2cm}\\
t_{\rec,3} : & \quad \text{when } T_\cl = T_\rec
\end{array}
\end{equation*}
$t_{\rec,1}$ indicates the time when the gas at $\tau=1$ starts to recombine and the recombination shell is formed. However, $\eta=1$ might still reside outside the recombination shell profile and is therefore not yet affected by recombination. At $t=t_{\rec,2}$, $\eta=1$ reaches the inner boundary of the recombination shell and $T_\cl=T_{\rsin}$ . The ability of the gas to thermalize is now impaired due to the steep drop in opacity inside the recombination shell. According to Eq \eqref{eq:delta}, the profile of $T_\gas$ inside the recombination shell behaves differently for $T_\gas<T_\rec$ and $T_\gas>T_\rec$. Therefore, the evolution of $T_\cl$ is also expected to change once $T_\cl=T_\rec$. The time when this happens is denoted as $t_{\rec,3}$.

Using the above relations and solving Equation set \eqref{eq:general}  for $m_{\rs}$, $T_{\cl}$ and $T_{\gas}$ together with Eq \eqref{eq:xion} for $x_\ion$ we find
\begin{flalign*}
& m_{\rs}(t)=m_{\cs} (t) =&
\end{flalign*}
\begin{equation}
 \left\{
  \begin{array}{l l}
m_{\cs}(t_{\rec,2})\left( \frac{t}{t_{\rec,2}}\right)
^{\frac{-(k-3)(63-33s-26k+11ks)}{(k-2)(9+2k)}} &  
    \quad, t_{\rec,2} \leq t \leq t_\L
     \vspace{0.5cm} \\
 m_{\rs}(t_\L)
 \left( \frac{t}{t_{L}}\right)^
 {\frac{52(k-3)}{29-33s+4k+11ks}} &    
   \quad, t \geq t_\L
  \end{array} \right.
  \label{eq:mrec}
\end{equation}
$$
\propto
 \left\{
  \begin{array}{l l}
     t^{2.26} &\quad, t_{\rec,2} \leq t \leq t_\L 
     \vspace{0.5cm} \\
     t^{1.89} &\quad, t \geq t_\L
  \end{array} \right. .
$$
where the specific values at $t>t_\reci$ are given for $k=6$ and $s=0.9$. As defined in Section \ref{sec:shell_propagation}, $t_\L$ is the time in which $m_\ls=m_\rs$ and the luminosity starts being affected by recombination. Indeed, the recombination shell, which at $t_{\rec,2}$ satisfies $m_{\rs}<m_{\ls}$, is moving inwards (in the Lagrangian sense) faster than the ionized luminosity shell (Eq \ref{eq:lum_old}), bringing the two shells together at $t_\L$. Note that the dependence of the temporal evolution of $m_{\cs}$ on the value of $k$ is rather strong. For example $m_{\cs}(t_{\rec,2} \leq t \leq t_\L) \propto t^{4.4}$, for $k=12$ and $s=0.9$. 

The gas temperature at $\tau=1$ is given by:
\begin{flalign*}
& T_{\gas,\tau=1} (t) =&
\end{flalign*}
\begin{equation}
 \left\{
  \begin{array}{l l}
T_{\gas, \tau=1}(t_{\rec,1}) \left( \frac{t}{t_{\rec,1}}\right)
 ^{-\frac{(k-3)(3-2k+ks-2)}{(k-2)(9+2k)}} &  
    \quad, t_{\rec,1} \leq t \leq t_\L
     \vspace{0.5cm} \\
 T_{\gas,\tau=1}(t_\L) 
 \left( \frac{t}{t_{L}}\right)
 ^{-\frac{2(167-147s-72k+49ks)}{31(29-33s+4k+11ks)}} &    
   \quad, t \geq t_\L
  \end{array} \right.
  \label{eq:Tgas}
\end{equation}
$$
\propto
 \left\{
  \begin{array}{l l}
    \propto t^{-0.15} &\quad, t_{\rec,1} \leq t \leq t_\L 
     \vspace{0.5cm} \\
    \propto t^{-0.10} &\quad, t \geq t_\L
  \end{array} \right. .
$$
The color temperature is:
\begin{flalign*}
& T_{\cl} (t) =&
\end{flalign*}
\begin{equation}
 \left\{
  \begin{array}{l l}
   T_{\cl}(t_{\rec,2}) \left( \frac{t}{t_{\rec,2}}\right)^{-\frac{2(k-3)(42-8s-36k+19ks)}{9(k-2)(9+2k)}} &
   \quad, t_{\rec,2} \leq t \leq t_{\rec,3}
  \vspace{0.5cm}\\
T_{\cl}(t_{\rec,3}) \left( \frac{t}{t_{\rec,3}}\right)
 ^{\frac{2(k-3)(42-8s-36k+19ks)}{31(k-2)(9+2k)}} &  
    \quad, t_{\rec,3} \leq t \leq t_\L
     \vspace{0.5cm} \\
 T_{\cl}(t_\L) 
 \left( \frac{t}{t_{L}}\right)
 ^{-\frac{2(167-147s-72k+49ks)}{31(29-33s+4k+11ks)}} &    
   \quad, t \geq t_\L
  \end{array} \right.
  \label{eq:Tcl}
\end{equation}
$$
\propto
 \left\{
  \begin{array}{l l}
    t^{-0.62} &\quad  , t_{\rec,2} \leq t \leq t_{\rec,3}
  \vspace{0.5cm} \\
     t^{-0.18} &\quad, t_{\rec,3} \leq t \leq t_\L 
     \vspace{0.5cm} \\
     t^{-0.10} &\quad, t \geq t_\L
  \end{array} \right. 
$$
where we assumed that at $t_\L<t$, $T_\cl<T_\rec$. When $m_{\rs}=m_{\cl}$ and before $T_{\cl}=T_{\rec}$, $T_{\cl}$ transitions into a faster decline than before as it propagates through the growing steepness of $T_{\gas}$ inside the recombination shell. Eventually, $T_{\cl}$ reaches $T_{\rec}$ at $t = t_{\rec,3}$. At this stage, the location of $\eta=1$ in the recombination shell is external the place where $T_{\gas}=T_{\rec}$ and the temperature evolves more slowly compared to its earlier evolution (Eq \ref{eq:dawn_cl}). Once $m_\rs = m_\ls$, $T_\cl$ evolves even slower. The dependence of the evolution of $T_{\cl}$ on the exact value of $k$ and $s$ both before and after $t_\L$ is very weak. 

Finally, the bolometric luminosity, defined by equation \eqref{eq:num_lum} is 
\begin{flalign*}
& L (t) =&
\end{flalign*}
\begin{equation}
 \left\{
  \begin{array}{l l}
L(t_{\rec,1}) \left( \frac{t}{t_{\rec,1}} \right)^{-\frac{2(1-s)(k-3)}{k-2}} &  
    \quad, t_{\rec,1} \leq t \leq t_\L
     \vspace{0.5cm} \\
 L(t_\L)  \left( \frac{t}{t_{L}}\right)
 ^{-\frac{6+8k+90s-30ks}{29+4k-33s+11ks}} &    
   \quad, t \geq t_\L
  \end{array} \right. 
  \label{eq:lum_mrec}
\end{equation}
$$
\propto
 \left\{
  \begin{array}{l l}
    \propto t^{-0.15} &\quad, t_{\rec,1} \leq t \leq t_\L 
     \vspace{0.5cm} \\
    \propto t^{0.32} &\quad, t \geq t_\L
  \end{array} \right. .
$$ 
At $t<t_\L$  equations \eqref{eq:lum_mrec} and \eqref{eq:lum_old} are similar since recombination does not affect the luminosity at this time. It depends more strongly on the value of $s$ and weakly on the value of $k$.
The increase (or slower decay for other values of $k$ and $s$) in the luminosity for $t>t_\L$ is expected as recombination increases the rate at which inner shells are exposed and release their internal energy. For reasonable values of $k$ and $s$, $L$ can be either rising slowly or decaying slowly with time.   

Using an outer profile of $k=12~;~s=0.9$ and an inner profile of $k=6~;~s=0.9$,  we find $t_{\rec,1}$, $t_{\rec,2}$ and $t_{\rec,3}$: 
\begin{equation} \label{eq:trec_RSG} 
 t_{\rec,1} \approx 2 \text{ days }  ~E_{51}^{0.07} M_{15}^{-0.12} R_{500}^{0.55}~ T_{\rec,7}^{-2.20} \kappa_{34}^{-0.60}  ,
\end{equation} 

\begin{equation} \label{eq:trec_RSG} 
 t_{\rec,2} \approx 4 \text{ days }  ~E_{51}^{0.14} M_{15}^{-0.18} R_{500}^{0.61}~T_{\rec,7}^{-1.98} \kappa_{34}^{-0.39}  ,
\end{equation} 

\begin{equation} \label{eq:trec_RSG} 
 t_{\rec,3} \approx 10 \text{ days }  ~E_{51}^{0.43} M_{15}^{-0.96} R_{500}^{0.90} ~T_{\rec,7}^{-2.35} \kappa_{34}^{-0.41}  ,
\end{equation} 
where $M_{15}$ is the mass of the progenitor in units of $15 M_{\odot}$, $R_{500}$  is its initial radius in units of $500 R_{\odot}$, $E_{51}$ is the explosion energy in units of $10^{51} \text{ erg}$, $T_{\rec,7}$ is the recombination temperature in units of $7,000$ K and $\kappa_{34}$ is the opacity in units of $0.34$ cm$^{2}$ g$^{-1}$. We also find the time at which recombination starts affecting the luminosity:
\begin{equation}
t_\L \approx 800 ~\text{days} ~ E_{51}^{0.08} M_{15}^{-0.61} R_{500}^{1.03}~ T_{\rec,7}^{-4.13} \kappa_{34}^{-1.57}  ~.
 \label{eq:tL_RSG} 
\end{equation}
As we discussed above, at $t_{\rec,2}$ only the observed temperature is affected by recombination, while the luminosity remains unaffected. Eq \eqref{eq:tL_RSG} implies that recombination does not necessarily affect the luminosity, since the entire energy might leak out of the envelope before the recombination shell has reached the luminosity shell. However, $t_\L$ is highly sensitive to the value of $k$. For example, for $k=9, s=0.9$ in the internal envelope, $t_\L \sim 40  \text{ days }$. Therefore, depending on the exact density profile of the envelope and on the progenitor properties, the luminosity can also be affected by recombination in some SNe.

The bolometric luminosity is given by 
\begin{flalign*}
& L(t) \approx &
\end{flalign*}
\begin{equation} \label{eq:lum_RSG}
   \left\{
  \begin{array}{l l}
   10^{42} \text{ erg s}^{-1} \times  \vspace{0.2cm}\\  ~E_{51}^{0.95} M_{15}^{-0.86} R_{500}~ \kappa_{34}^{-0.91} t_{\text{day}}^{-0.15}
     &  ,t_{\rec,1}<t<t_\L\\
     \\10^{41} \text{ erg s}^{-1} \times  \vspace{0.2cm}\\  ~E_{51}^{0.92} M_{15}^{-0.57} R_{500}^{0.51} T_{\rec,7}^{1.97}~ \kappa_{34}^{-0.18}  t_{\text{day}}^{0.32}
    &    ,t>t_\L\\
  \end{array} \right. .
\end{equation}
The observed temperature changes its evolution with time from $t_{\rec,2}$, 
\begin{flalign*}
& T_{\cl}(t) \approx&
\end{flalign*}
\begin{equation} \label{eq:temp_RSG}
    \left\{
      \begin{array}{l r}
   20,000 \text{ K}  \times  \vspace{0.2cm}\\ ~E_{51}^{0.27} M_{15}^{-0.41} R_{500}^{0.56} T_{\rec,7}^{-0.47} \kappa_{34}^{-0.25}   t_{\text{day}}^{-0.62}
       &  t_{\rec,2}<t<t_{\rec,3}\\
   \\10,000 \text{ K} \times \vspace{0.2cm}\\  ~E_{51}^{0.08} M_{15}^{-0.12} R_{500}^{0.16} T_{\rec,7}^{0.57} \kappa_{34}^{-0.07}  t_{\text{day}}^{-0.18}
       &  t_{\rec,3}<t<t_\L\\
   \\7,000 \text{ K} \times \vspace{0.2cm}\\ ~E_{51}^{0.07} M_{15}^{-0.07} R_{500}^{0.08} T_{\rec,7}^{0.89} \kappa_{34}^{0.05} t_{\text{day}}^{-0.10}
    &  t>t_\L \vspace{0.2cm}\\
  \end{array} \right.
\end{equation}
After $T_\cl$ reaches $T_\rec$ (at $t = t_{\rec,3}$), the observed temperature depends very weakly on the progenitor and explosion parameters but is naturally more strongly affected by the value of $T_{\rec}$. After reaching $T_{\rec}$, $T_{\cl}$ evolves very slowly with time: at $t_{\rec,3}$ it is $\approx 7,000$ K and reaches $\approx 5,000$ K by the end of the plateau, around $t\sim100$ d.

\begin{figure}
\includegraphics[width = \columnwidth]{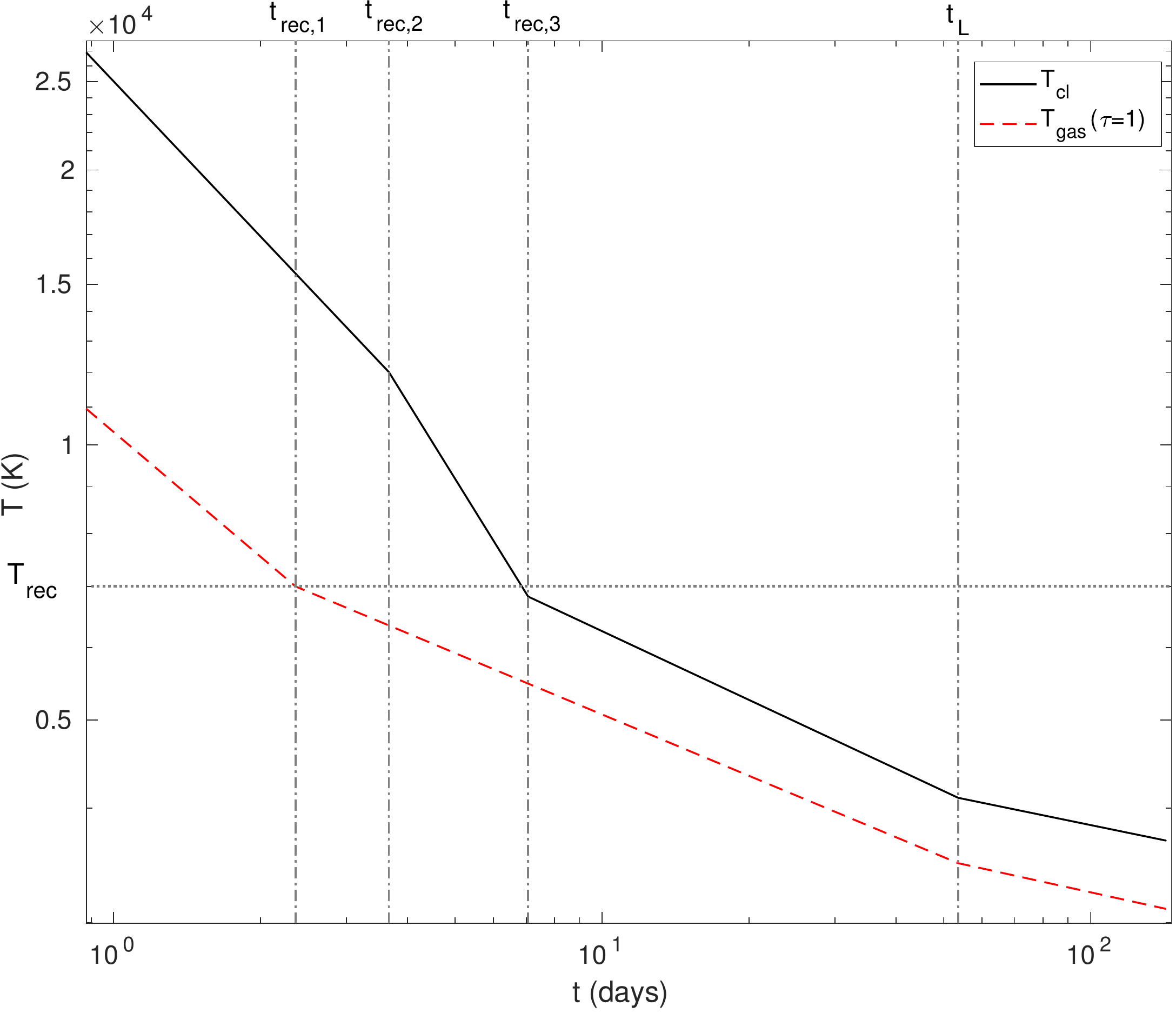}
\caption{Solid line: the evolution of $T_{\cl}$ according to Eq \eqref{eq:temp_RSG}. Dashed red line: the evolution of $T_{\gas}$ at the outer boundary of the recombination shell ($\tau=1$) according to Eq \eqref{eq:Tgas}. We used the following parameters for the progenitor: $k=8$, $s=0.9$, $E_{51}=1$, $M_{15}=1$, $R_{500}=1$ and $T_{\rec,7}=1$. The dash-dotted lines indicate the locations of $t_{\rec,1}$, when the gas at $\tau=1$ starts to recombine, $t_{\rec,2}$, when $T_\cl=T_{\rs,in}$ and $T_\cl$ is first affected by recombination, $t_{\rec,3}$, when $T_\cl=T_\rec$ and begins to decline slowly, and $t_\L$, when the recombination wave reaches the luminosity shell.} \label{f:temp_analytic}
\end{figure}

In Figure \ref{f:temp_analytic} we plot $T_{\cl}$ (black solid line) together with $T_{\gas}(\tau=1)$ (red dashed line) obtained for an RSG with $E_{51}=1$, $M_{15}=1$, $R_{500}=1$ and $T_{\rec,7}=1$, and the hydrodynamic parameters $k=12,s=0.9$ in the external envelope and $k=8,s=0.9$ in the internal envelope. We use a steeper density profile in order to demonstrate the effect of recombination on $T_\cl$ at $t>t_\L$.
Before recombination reaches the colour shell, $T_{\cl}$ decreases rapidly. As soon as $T_\cl = T_{\rs-in}$, $T_{\cl}$ drops more rapidly until reaching $T_{\rec}$, and then transitions to a slower evolution, keeping a roughly constant temperature around $T_{\rec}$. For these particular parameters of the progenitor, $t_\L$ is reached before the whole envelope is exposed, as $t_\L\approx 50$ d. At $t>t_\L$ the temperature evolves even more slowly.

\subsection{The single band light curves and the `Plateau'}
\begin{figure}
\includegraphics[width = \columnwidth]{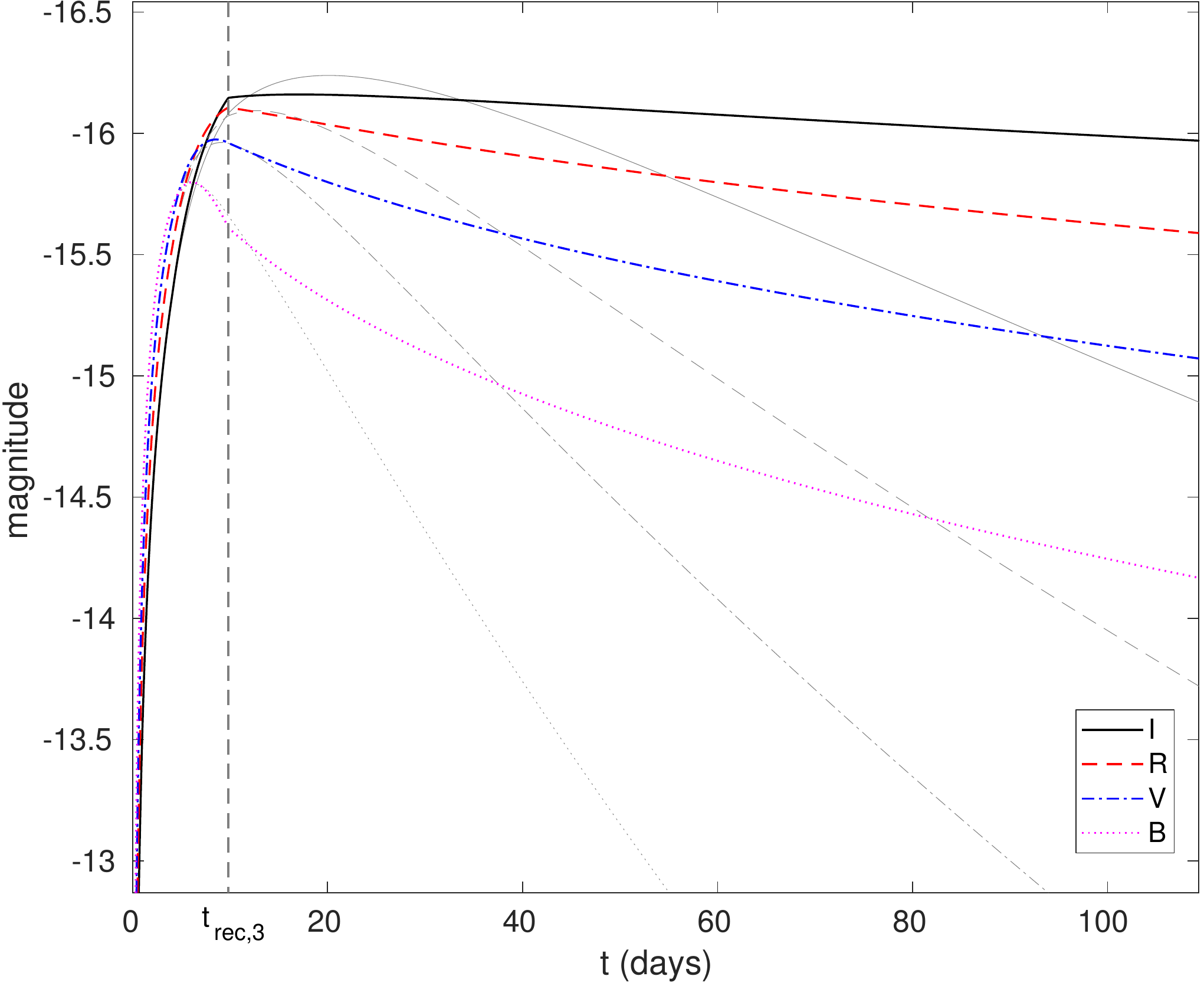}
\caption{Single-band light curves constructed from the luminosity and temperature profiles described in Equations \ref{eq:lum_RSG} and \ref{eq:temp_RSG} for the following parameters: $k=6$, $s=0.9$, $E_{51}=1$, $M_{15}=1$. $R_{500}=1$, $\kappa_{34}=1$ and $T_{\rec,7}=1$.} \label{f:plateau_magnitudes}
\end{figure}

Although $L$ remains unaffected by recombination throughout most (or all) of the photospheric evoluion, the single band light curves, which are the SN observables, are highly affected by the flattening of $T_{\cl}$. In Figure \ref{f:plateau_magnitudes} we plot the light curves in $B$ ($4450 {\AA}$), $V$ ($5510 \AA$), $R$ ($6580 \AA$) and $I$ ($8060 \AA$) bands resulting from the analysis in Section \ref{sec:analytic_LC} for a progenitor with the parameters $E_{51}=1$, $M_{15}=1$, $R_{500}=1$, $\kappa_{34}=1$ and $T_{\rec,7}=1$, and a density profile with $k=12$, $s=0.9$ in the outer enveople and $k=6$, $s=0.9$ in the inner envelope. In light grey we plot the light curves that would have been observed had there been no recombination, namely, assuming the temperature continues to decline as it did before the gas started to recombine (Eq \ref{eq:dawn_cl}). Prior to the light curve peak, $T_{\cl}$ is high and the central frequency of the photometric band is lower than the peak frequency of the blackbody distribution. 
The luminosity in a certain frequency $\nu$ near the Rayleigh-Jeans tail satisfies:
\begin{equation}
L_{\nu} \propto \frac{r_\cs^2 T_{\cl}}{\tau_\cs} \propto \frac{L}{T_{\cl}^3} \propto t^{1.51}.
\end{equation}
The above equation implies that if $T_{\cl}^3$ decreases more quickly than $L$ (which is indeed the case), then the flux in $\nu$ will rise. Hence, the drop in temperature together with the relatively moderate decline in the luminosity increases the flux of a given frequency as the peak of the blackbody approaches the central band frequency. 
When there is no recombination (grey lines in Figure \ref{f:plateau_magnitudes}), the light curves decline moderatey after reaching the peak magnitude. However, as $T_\cl$ reaches $T_\rec$ (at $t\sim 10$ d in this case), it flattens, and since both $L$ and $T_{\cl}$ are roughly constant at this stage ($L$ declines slowly due to the hydrodynamic profile of the envelope, and not due to recombination), the flux in a specific band remains almost constant and the observed `plateau' is produced. The plateau is more pronounced in lower frequencies that reside farther from the spctral blackbody peak, where the flux is less sensitive to the change in temperature. This is why the plateau is flatter in the $I$-band than in the $V$-band. As an indication of how important recombination is to the plateau, we show in Figure \ref{f:plateau_magnitudes} that when recombination was ignored (grey curves), the light curves decline much faster.

\begin{figure}
\includegraphics[width = \columnwidth]{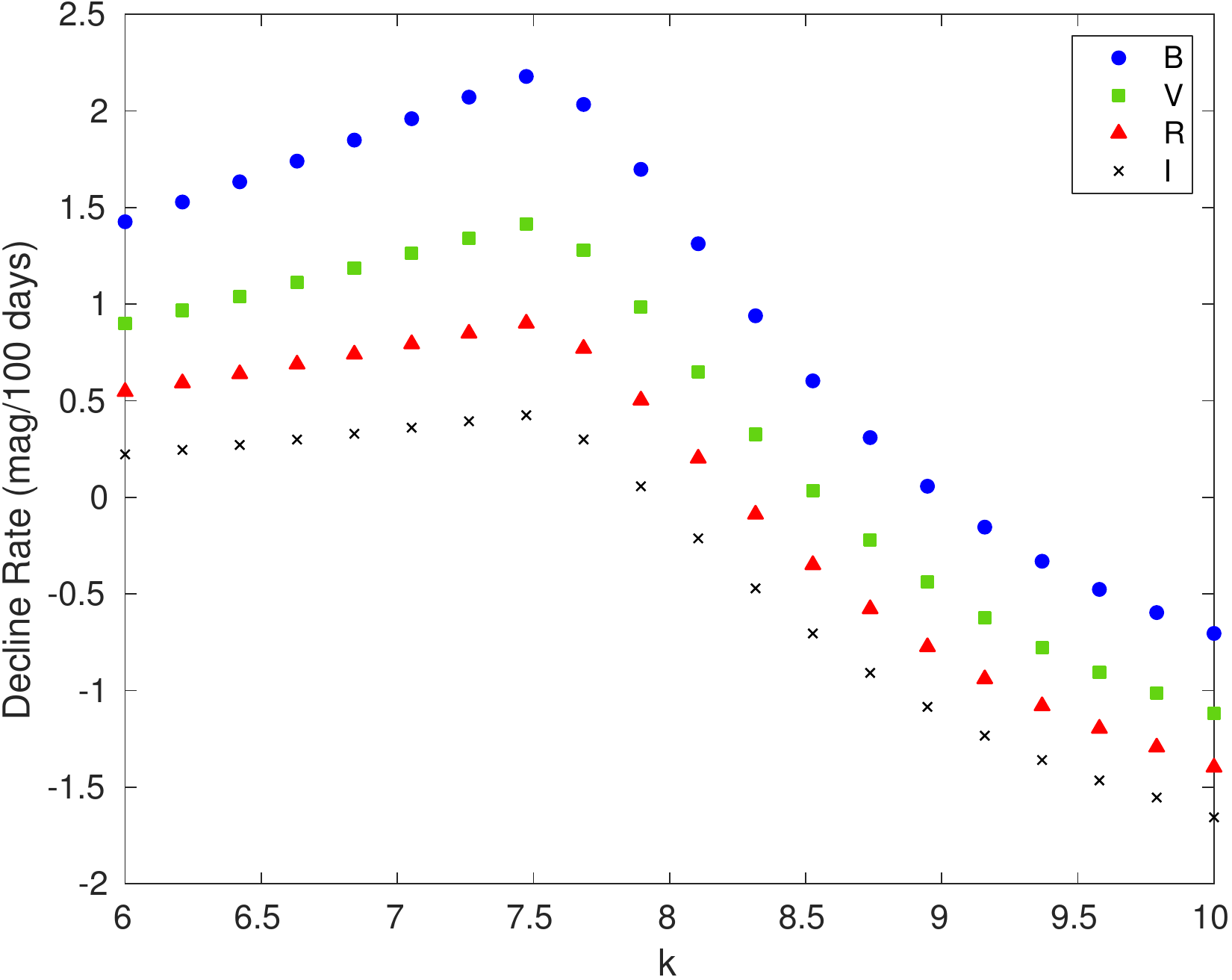}
\caption{The decline rates in different photometric bands as a function of the density power law index $k$, calculated between day 20 and day 80 after the SN explosion, for a progenitor with the parameters $E_{51}=1$, $M_{15}=1$, $R_{500}=1$, $\kappa_{34}=1$ and $T_{\rec,7}=1$. The decline rates increase as the density profiles become steeper, but reach a maximum around $k\sim 8$, as the recombination shell reaches the luminosity shell before day 80 and causes the luminosity to rise. } \label{f:LC_declinerates}
\end{figure}

We find that the decline rates of the single band light curves are most sensitive to the parameter $k$, where steeper density profiles produce more rapidly declining light curves, since $T_\cl$ decreases faster with $k$. However, when $k\gtrsim 7.5$, the time $t_\L$ is reached before the whole envelope is exposed, and causes the bolometric luminosity to rise. As a result, the decline rate of the light curves is expected to be lower due to the faster release of radiation. In Figure \ref{f:LC_declinerates} we plot the decline rates as a function of $k$ in the $B$, $V$, $R$ and $I$ bands for a progenitor with the parameters $E_{51}=1$, $M_{15}=1$, $R_{500}=1$, $\kappa_{34}=1$ and $T_{\rec,7}=1$, calculated by fitting a line to the light curve in the time range $20-80$d. The decline rate increases with $k$, but reaches a maximum due to the increase in luminosity at $t>t_\L$, which occurs earlier for higher values of $k$. The overall trend of the light curve can even be increasing, if the density profile is steep enough ($k\gtrsim 8.5$ in the $V-$band).
In the $V-$ band, the decline rates range from $\sim 2.5 \text{ mag}/100~\text{days}$ to $\sim -0.5 \text{ mag}/100~\text{days}$. These values are in agreement with the observed $V-$ band decline rates of type-II SNe published in \cite{Anderson14}, who find a range of $\sim 3.5 \text{ mag}/100~\text{days}$ to $\sim -0.8 \text{ mag}/100~\text{days}$. Thus, the distribution in density profiles can potentially account for some of the observed diversity in decline rates, and explain the difference between II-P and II-L SNe.

The plateau duration, $t_{\text{SN}}$, can be approximated as the time at which the energy deposited by the shock is released from the \textit{entire} envelope. 
We examine the energy release by the ``last" shell, constituting the bulk of the star. It has a mass $\sim M_{\text{ej}}$, radius $r=v_{\sn}t_{\sn}$
and density $\sim M_{\text{ej}}/r^3$, where $M_{\text{ej}} \sim M$ is the ejected mass and $v_{\sn} \approx \sqrt{2E/M_{\text{ej}}}$.
If recombination does not affect the luminosity (i.e., $t_\L>t_{\text{SN}}$), the relations derived by \citet{arnett80} are valid:
\begin{equation}
	\begin{array}{lll}
		t_{\text{SN}} &\propto&  E^{-0.25} M^{0.75} \kappa_\T^{1/2} \\
		&&\\
		L_{\text{SN}} &\propto& E~ M^{-1} R ~\kappa_\T^{-1}  
	\end{array}
\label{eq:lsn_Arnett}
\end{equation}
where $L_{\sn}$ is the typical luminosity of the plateau

When recombination affects the luminosity, the plateau luminosity and duration can be determined by the following relation
$$
E_0 \left( \frac{r}{R_{\star}} \right)^{-1} t_{\sn}^{-1} \approx L_{\sn} \approx
\frac{a c T_\rsout^4 r^3} {x_{\ion}(T_\rsout) \kappa_\T \rho r^2} ,
$$
where $E_0 \approx E/2$ is the initial internal energy. We use the ionization model according to Eq \eqref{eq:xion} 
where $T_\rsout$ and $x_{\ion}(T_\rsout)$ are found by the condition at the external boundary of the recombination shell ($\tau=1$, Eq \ref{eq:num_tau}).
Using the above definitions, we find for $t_{\text{SN}}>t_\L$:
\begin{equation}
	\begin{array}{lll}
		t_{\text{SN}} &\propto& E^{-0.18} M^{0.47} R^{0.21} T_{\rec}^{-0.85} \kappa_\T^{0.07} \\
		&&\\  
		L_{\text{SN}} &\propto& E^{0.86} M^{-0.44} R^{0.58}T_{\rec}^{1.69}\kappa_\T^{-0.15}  
	\end{array}
\label{eq:ltsn}
\end{equation}

Therefore, if $E,R$ and $M$ are independent, the variation in $t_{\text{SN}}$ is expected to be larger than the observed, almost constant, plateau duration of 100 days \cite[]{arcavi12}. This implies that there is probably a dependence between these explosion properties \citep{Poznanski13}.

\subsection{Semi-analytic light curve for a numerical hydrodynamic profile} \label{sec:semi_analytic}
Equation set \eqref{eq:general} can be solved semi-analytically for any $x_{\ion}(\rho,T)$,  $\rho(m)$, $E_{\ad}(m)$  and $v(m)$. This enables us to take a hydrodynamic profile that was calculated using a hydrodynamic simulation (without any radiation transfer) and derive the light curve it produces from the shock breakout and up to the end of the plateau.

Here we present the light curves derived for a numerical profile of an RSG explosion calculated by Roni Waldman (private communication). Since we do not take the effect of $^{56}$Ni into account, we use a progenitor with $M_{\Ni} = 0$. The progenitor had a zero age mass of 15$M_\odot$ and was evolved numerically using the MESA STAR code up to explosion. Its pre-explosion radius is $R=630 R_{\odot}$, the ejecta mass is $13M_{\odot}$ and the explosion energy is $E=1.0\times 10^{51}$ erg. By fitting the density and the internal energy profiles, we find that $s=0.85$ throughout the envelope, and $k=12, k=9$ in the external and internal regions, respectively.
We take the profiles $\rho_i=\rho(m,t_i)$, $E_{\ad,i}=E_{\ad}(m,t_i)$, $r_i=r(m,t_i)$ and $v(m)$ at some early time $t_i$ (before recombination starts) and consider all the shells that already reached their coasting velocity (the spherical phase).
We define the optical depth $\tau_i$ and the diffusion time $t_{d,i}$ of each shell $i$:
\begin{subequations} \label{eq:tau_td_sa}
\begin{equation}\label{eq:tau_sa}
\tau_i(m)=\int_{r_i}^\infty \kappa_\T x_{\ion}(T_{\gas}) \rho dr^{\prime}
\end{equation}
\begin{equation}\label{eq:td_sa}
t_{d,i}(m)=\int_{r_i}^\infty \frac{3\tau_i}{c} dr^{\prime}
\end{equation}
\end{subequations}
Where for $t=t_i$, $x_{\ion}=1$. We then construct the profiles of $r$, $\rho$ and $E_{\ad}$ at later times, $t>t_i$ :
\begin{subequations}\label{eq:evolved_parameters}
\begin{equation}
r(m,t)= r_i \left( \frac{t}{t_i} \right)
\end{equation}
\begin{equation}
\rho(m,t)=\rho_i \left( \frac{t}{t_i} \right)^{-3}
\end{equation}
\begin{equation}
E_{\ad}(m,t)=E_{\ad,i} \left( \frac{\rho}{\rho_i} \right)^{\gamma-1} ~.
\end{equation}
\end{subequations}
We verify that throughout the evolution the energy in the envelope is radiation dominated, and use an adiabatic index of $\gamma = 4/3$. The temporal evolution of $\tau$ and $t_\d$ changes as recombination starts in the envelope and $x_{\ion}$ drops. In order to compute the ionization fraction of the gas at $t>t_i$ we find the profile of $T_{\gas}$ throughout envelope. According to Eq \eqref{eq:e_Tg} and Eq \eqref{eq:eps}, $T_\gas$ is given by $T_\gas = \left(L \tau /4 \pi \sigma r^2\right)^{1/4}$, where $L$ is determined at the place where $\tau=c/v$. It can be found easily for $\tau<1$. However, since $\tau$ is itself a function of $T_\gas$, for $\tau>1$, $T_\gas$ is found by propagating the physical properties at $\tau=1$ inwards. Then, we solve equation set \eqref{eq:general}, for $m_\lum(t)$, $m_\cl(t)$, $m_{\rs}(t)$, $T_{\gas, \tau = 1}(t)$ and $T_{\cl}(t)$.

\begin{figure}
\includegraphics[width=1\columnwidth]{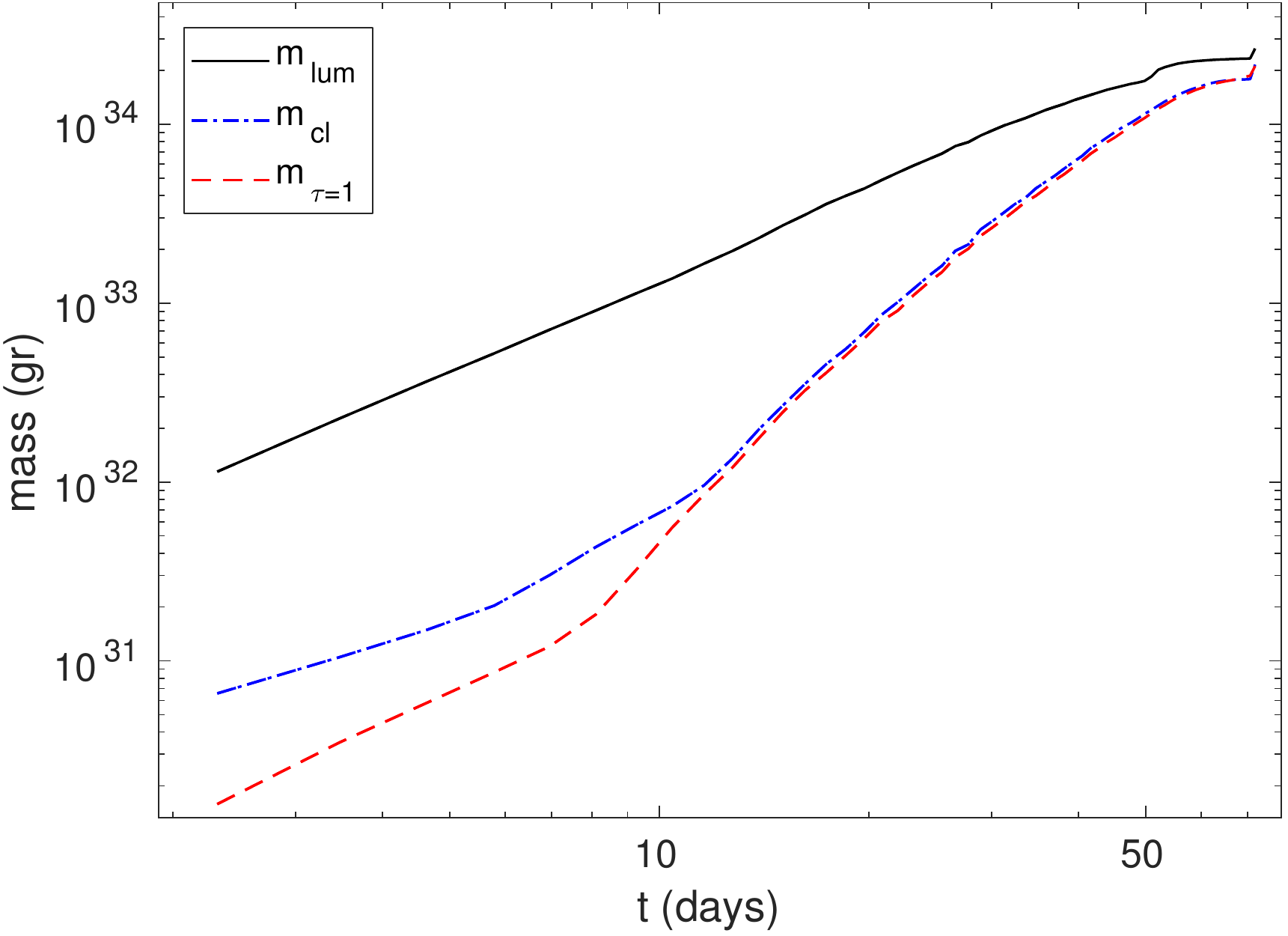}
\caption{The evolution of the three shells with time. When the gas at $\tau=1$ stars recombining the recombination shell moves inwards faster than the color shell ($\sim 8d$), and the locations of $\tau=1$ and $\eta=1$ coincide at $\sim 11d$. The two shells propagate together and reach the luminosity shell at $\sim 40d$. The internal energy in the envelope is then released quickly and all the energy has escaped the envelope by day $\sim 60$.} \label{f:shells}
\end{figure}

\begin{figure}
\includegraphics[width=1\columnwidth]{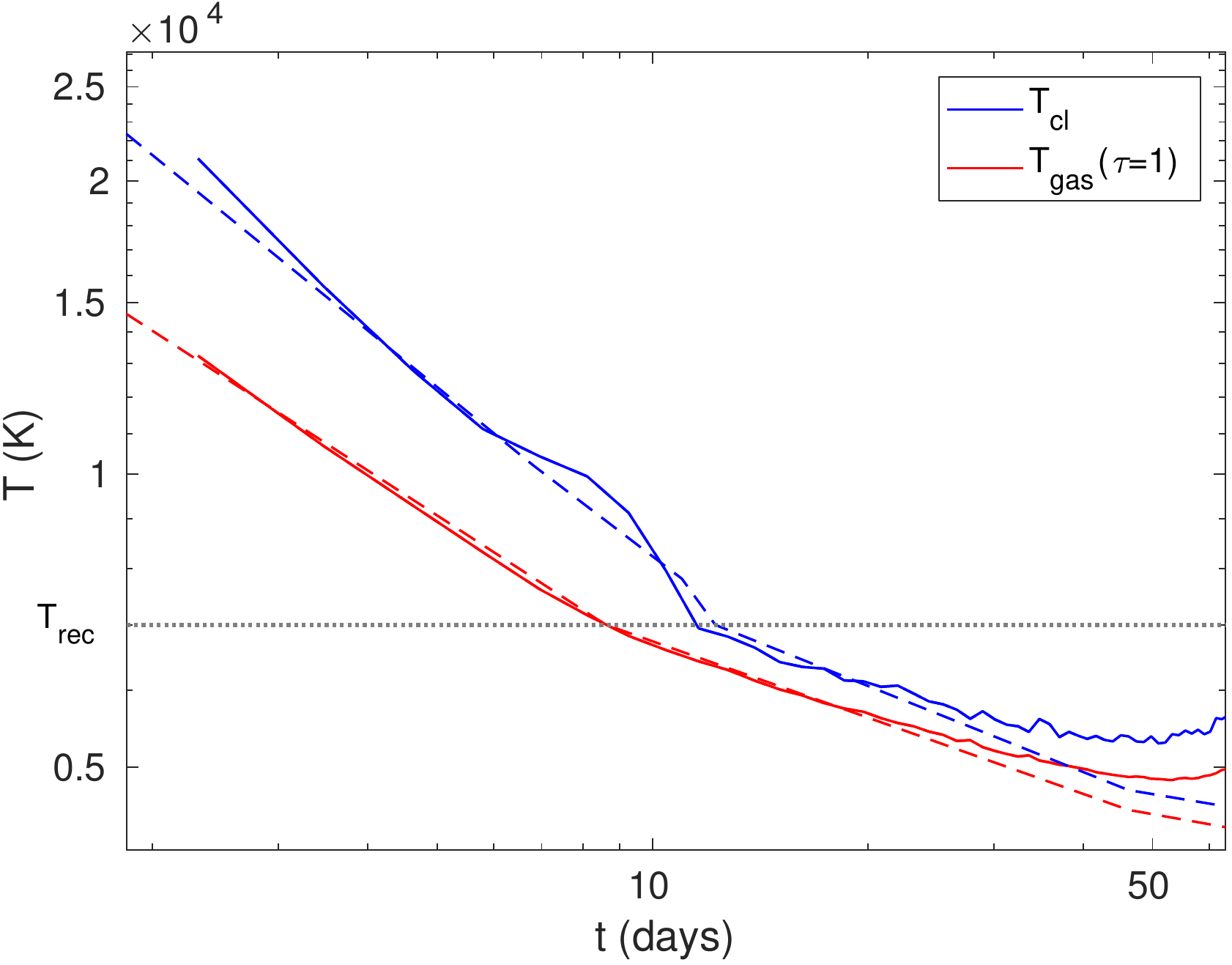}
\caption{The evolution of $T_{\cl}$ (solid blue line) and $T_{\gas}(\tau=1)$ (solid red line) with time, resulting from the semi-analytic calculation. The analytic predictions for the progenitor parameters are plotted as the dashed lines according to Eq \eqref{eq:Tcl}, where $T_\cl$ was normalized to match the semi analytic result at $t\sim 1$ d . When $\eta=1$ is first satisfied inside the recombination shell $T_\cl$ starts declining more rapidly and then flattens once $T_\cl=T_\rec$. } \label{f:T_num}
\end{figure}

The calculation was performed up to the point that the luminosity shell ends crossing the H envelope. In this specific model, the recombination shell is expected to arrive at the luminosity shell before all the energy escaped from the envelope at $t\sim 43$ d. In Figure \ref{f:shells} we plot the evolution of $m_\lum$, $m_\rec$ and $m_\cl$ with time. Before day $\sim 10$ the three shells evolve independently. Shortly after the gas temperature at $\tau=1$ reaches $T_{\rec}$, the recombination shell reaches the color shell and they start evolving together, as seen in Figure \ref{f:shells}. According to Eq \eqref{eq:Tcl}, $T_{\cl}$ is predicted to decrease more rapidly than before, until reaching $T_{\rec}$. When $T_{\cl} = T_{\rec}$ the color temperature is expected to flatten and remain almost constant near $T_{\rec}$. This evolution is clearly evident in Figure \ref{f:T_num}, where the theoretical predictions are the dashed lines. To fit the semi analytic results, the normalization of the analytic solution of $T_\cl$ at $t=1$ day was multiplied by $1.2$.

\begin{figure}
\includegraphics[width=1\columnwidth]{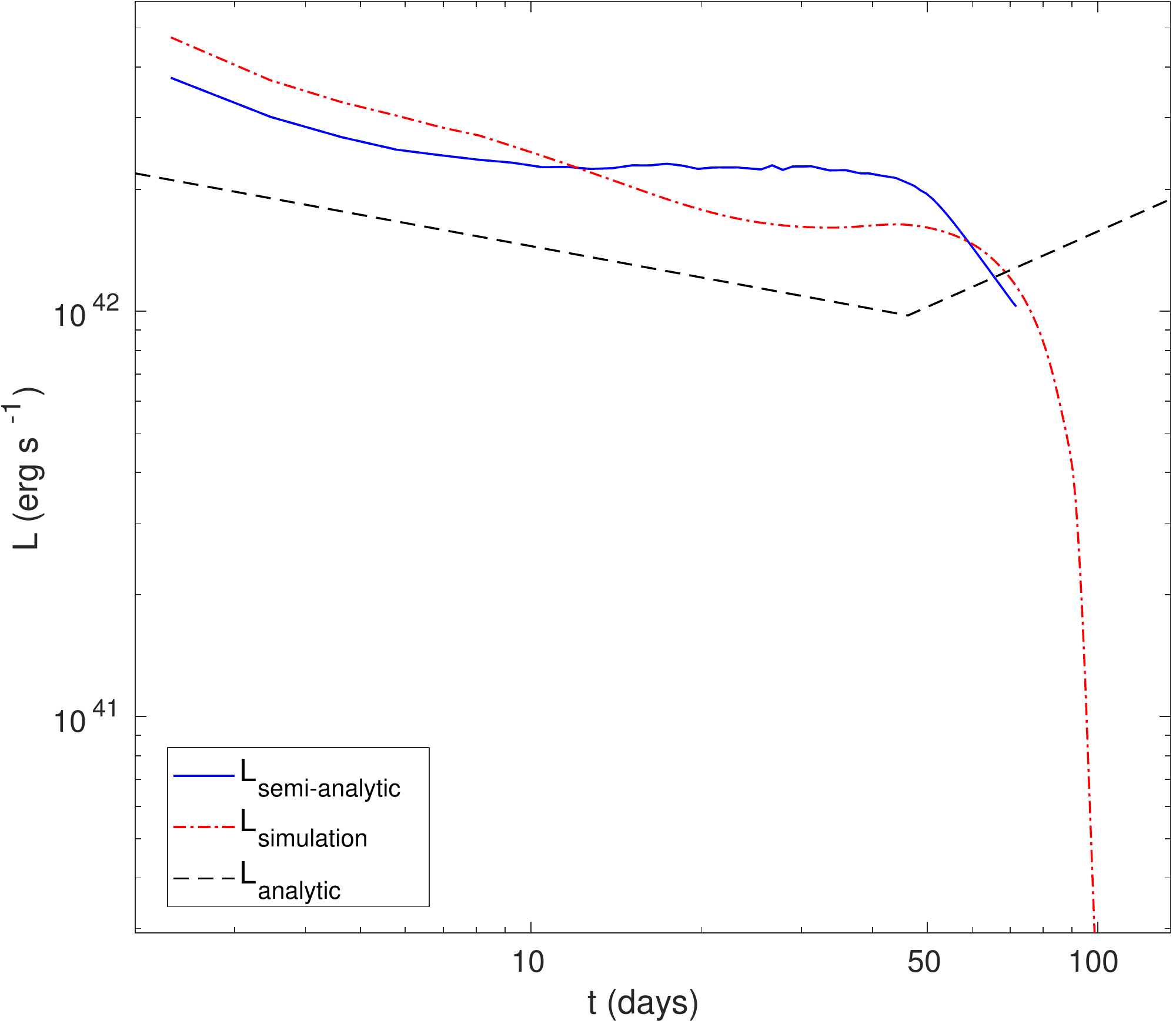}
\caption{The evolution of the bolometric luminosity from the semi-analytic solution (solid blue line), the analytic prediction (black dashed line) and the result from the radiative transfer simulation using the code Vulcan. Around day $\sim 40$, the recombination shell reaches the luminosity shell and the release of energy becomes dominated by recombination instead of diffusion, causing the luminosity to increase. This effect is observed (though more subtly) in the numerical light curve.} \label{f:L_num}
\end{figure}

In Figure \ref{f:L_num} we plot the luminosity calculated with the semi-analytic solution (solid blue line), and the analytic prediction (dashed black line). In our analytic description, a significant effect on $L_\bol$ occurs when the recombination shell reaches the luminosity shell. However, in the semi-analytic calculation we calculate $\tau$ by integrating over radii external to $r$ (Eq \ref{eq:tau_sa}), and therefore the location of the luminosity shell is mildly affected even before $t=t_\L$ (at $t = 15$ d, the change in $m_\ls$ is $\sim 20\%$). The luminosity shell is constantly "pushed" inside, and as a result the recombination shell reaches it only at late times (see Figure \ref{f:shells}). As seen in Figure \ref{f:L_num}, $L$ begins to flatten at $t\sim 10$ d and does not show the sharp re-brightening predicted by the analytic model.
We also include the light curve calculated with the radiative transfer code Vulcan \citep{Livne93} for this progenitor. The numerical light curve shows a similar behaviour to our semi-analytic solution, and transitions into a flat evolution at $t\sim 25$ d, later than our semi-analytic solution. The reason is that $T_\gas$ is treated more accurately in Vulcan through the equation of state, and the onset recombination is at $\sim 20$ d. Flattening or re-brightening of the bolometric luminosity at $t\sim 20-30$ d is observed in various type-II SNe (see Figure 5 in \citealt{Faran18}), in accordance with the semi-analytic calculation.

\section{Comparison to Previous Works} \label{sec:comparison}

\subsection{Analytical Works}

Several works have dealt with the problem of recombination in SN envelopes (e.g. Grassberg \& Nadyozhin 1975, Popov 1993). These works examined a recombination wave that propagates inwards through the hydrogen envelope and dictates the energy emission\footnote{
Some works use the term ``wave of cooling and recombination'' (WCR) or ``cooling and recombination wave'' (CRW).}. We compare the assumptions of some of those works and their results to our model.

\citet{grass_nad76} examined a recombination wave propagating in a SN envelope. Here we briefly summarize their results, focusing on the recombination shell and its boundary conditions. They consider a homologously expanding envelope with a density profile of $\rho \propto r^{-k} t^{k-3}$, similar to our model. Inside the envelope, they assume the existence of a `cooling front (CW)', which divides the envelope into two regions - a hot region internal to the CW, and a cool region external to the CW. The opacity near the CW is considered to be dominated by electron scattering, with $\kappa \approx 0.4 x_{\ion}$, where $x_{\ion}$ is found according to the SAHA equation. \citet{grass_nad76} look at the the conditions in the internal and external boundary of the CW. As opposed to what we assume in this work, they give a significant role to the energy released from the recombination itself in determining the temperature at the inner region of the recombination wave,  $T_1$. In their model $T_1$, is determined by the balance between adiabatic cooling and heating by recombination photons. The temperature at the external boundary of the CW, $T_2$, is given by the condition that the optical depth at the outer boundary satisfies $\tau_{\rs,out}=1$. The color temperature is assumed to be $T_2$, which is the temperature at the photosphere. Therefore \citet{grass_nad76} do not consider the thermal coupling between the radiation and the gas. 
The energy per unit mass in a shell, neglecting the matter internal energy, is assumed to be 
$$E \approx \frac{aT^4}{\rho} +  \frac{\chi \rho_{\rs}}{m_{\mu}} ,$$
where $\chi$ is the recombination potential and $m_{\mu}$ is the average mass per ion. The importance of each contribution is density dependent, where the recombination energy dominates at high densities.
In their model, the difference in the internal specific energy between the inner and outer boundaries, $\Delta E = E_1 - E_2$, dictates the propagation velocity of the CW. 

When the CW reaches deep enough regions, the energy in the internal boundary of the recombination shell is dominated by recombination and $\Delta E \approx E_1 \approx \chi \rho_{\rs}/m_{\mu}$. The diffusion time in the external boundary
is $\sim r_{\rs}/c$. Equating the luminosity from the external boundary to
the release of the energy in the internal boundary during the dynamical time of the shell ($\sim t$)
gives 
$$\rho_{\rs}/t \propto T_2^4/r_{\rs}.$$
Under the assumption that $T_2\approx$ constant, one finds that $r_{\rs} \propto t^{\frac{k-4}{k-1}}$.
Since in this analysis $L \propto r_{\rs}^2$, then $L \propto t^{0.8}$ for $k=6$. 
In contrast, we find that the energy from the recombination itself does not play any significant role in
determining the dynamics of the recombination shell. Let us consider late times, when $m_{\rs}=m_{\lum}$.
The luminosity released from the internal boundary of the recombination shell, including the energy from recombination, is
$L=E_{\ad}(m_{\rs})/t+\chi \rho_{\rs} r_{\rs}^3 /m_{\mu}t$. Thus, the recombination energy is important
when 
$$\frac{\chi}{k_B T_{\bb,\rs-\in}} \beta_{\rs-\in} \gg 1,$$
where $\beta_{\rs-\in}$ is the ratio between gas pressure and radiation pressure in the internal boundary of the shell:
$$\beta_{\rs-\in} \sim \frac{k_B T_{\rs-\in} \rho_{\rs}}{a T_{\rs-\in}^4 m_{\mu}} \sim 10^{-2} 
\left( \frac{k_B T_{\rs-\in}}{\text{eV}} \right) ^{-3}.$$
We find that this condition is satisfied only if $T_{\rs-\in} < T_{\rec}$. Since in our model $T=T_{\rec}$ occurs where $\delta(T_{\gas})$ inside the recombination shell is minimal, and this happens well within the recombination shell, this condition cannot hold. Therefore, the assumption that the recombination energy affects the evolution of the recombination shell is not justified. A simpler argument is that the total energy released when $10~M_\sun$ of H recombine is $10^{47}$ erg, which is $\sim2$ orders of magnitude lower than the total energy released by type-II SNe.

\cite{Chugai90} used a simple model for recombination where a cooling wave of zero width divides the envelope into an internal opaque region and an external transparent region. Here neither the ionization fraction of the gas nor the coupling between radiation and gas were taken into account. The photospheric temperature is therefore constant and equal to the recombination temperature. \cite{Chugai90} solved for the radius of the cooling wave and finds a parametric relation for the plateau duration in the presence of recombination, which is the main result of his paper:
$$t_p^{Chugai 1991} \propto E^{-1/8} M^{3/8} R^{1/4} T_\rec ^{-1}.$$
Despite the simplicity of his model, the dependences \cite{Chugai90} found on the progenitor parameters are not very different from those we find in Eq \eqref{eq:ltsn}. \cite{Chugai90} points out that there is no dependence on the opacity of the envelope, which is comparable to the very weak dependence we find on $\kappa$ in Eq \eqref{eq:ltsn}.

\citet[]{popov93} considered a homogeneous density in the freely expanding envelope.
He used an analytic solution by \citet[]{arnett80} that ignores recombination ($\kappa=\kappa_\T$).
This solution sets conditions before recombination starts that are not realistic in the actual case of a non-homogeneous envelope. The model assumes a step function opacity, 

$$   \kappa = \kappa_\T \left\{
  \begin{array}{l l}
    1 & \quad   ,T>T_{\rec}\\
    0  & \quad ,T<T_{\rec}\\
  \end{array} \right. 
$$
and does not deal with color temperature and coupling between matter and radiation, so the onset of 
recombination is defined by the time when the temperature at the photosphere is $T_{\rec}$. This turns out to be the right condition, since recombination starts affecting the temperature when $T_{\gas}(\tau=1)=T_{\rec}$. However, since the opacity becomes $0$ when recombination starts, the color temperature is fixed on $T_\rec$ and only the bolometric luminosity is allowed to change. 
\cite{popov93} solved for the location of the cooling wave as in \cite{Chugai90} and found an expression for the plateau duration

$$t_p^{Popov 1993} \propto R^{1/6} M^{1/2} E^{-1/6} T_\rec ^{-2/3} \kappa^{1/6}.$$

This relation also resembles the dependence we found, except for the stronger dependence on the opacity.

\subsection{Observational works}
Several observational works have computed the temporal evolution of the color temperature and the luminosity in type-II SNe. Such works include \cite{Bersten09}, \cite{Lyman14}, \cite{Valenti16}, \cite{Lusk17} and \cite{Faran18}. The observed temperature curves show a clear flattening as they enter the range of the expected recombination temperatures in type-II SNe.

A detailed analysis of the decline rates of the color temperature and bolometric luminosity was carried out in \cite{Faran18}, who computed the color termpatures and logarithmic derivatives of 29 SNe by fitting black body models to their spectral energy distributions. They then calculated the logarithmic derivatives of the early and late evolution. The SNe in their sample showed a variety of early decline rates for the temperature around the expected values, but all of them transitioned into a much slower evolution when their color temperatures reached $\sim 7000K$, with power laws between the values $-0.05$ and $-0.14$. These values are in agreement with the theoretical decline rated predicted in this work. In most of the objects they also find a slower decline rate of the bolometric luminosity at late times compared to the early evolution.

\section{Summary and Discussion} \label{sec:summary}

We develop an analytic model that describes the implication of recombination on the evolution of SN light curves. Recombination sharlpy decreases the number of free electrons in the envelope and causes the opacity to drop dramatically. This directly affects the radiation-gas coupling via the ability of the gas to absorb and emit photons, which determines the radiation temperature.

In our model, we consider three shells that dominate the properties of the SN temperature and luminosity: (1) The luminosity shell -- the location in the envelope out of which photons are able to diffuse over a dynamical time, which is the source of the observed radiation. This shell determines the radiative flux through all the shells external to it. (2) The color shell -- the outermost shell within which photons that originated in the luminosity shell are able to thermalize with the gas. This is the shell where the color temperautre is determined. (3) The recombination shell -- the shell that contains the recombination front, i.e., the outermost shell which is fully ionized in its inner boundary and highly recombined in its outer boundary. When the recombination shell reaches the color shell, the two shells coincide and propagate together. At late times this shell may also coincide with the luminosity shell and govern the energy release.

The main advantage of our model compared to previous works is that we use the profile of the gas temperature inside the recombination shell to account for the coupling between radiation and gas. Since the color temperature is determined inside the recombination shell at $t>t_{\rec,2}$, the relation between the two (through the effect of the gas temperature on the ionization fraction) is crucial in finding the color temperature as a function of time. Our model provides more reliable estimates to the observed temperature and the bolometric luminosity, and offers new insights into the recombination process and the relations between the observables and the progenitor and explosion properties. 

We find that when recombination starts it first affects only the temperature, which becomes almost constant with time. The bolometric luminosity, at first, continues to drop very slowly without being affected by recombination. Only later the recombination wave reaches the luminosity shell. The temperature from that point evolves even more slowly, while the bolometric luminosity flattens and may even rise moderately. Whether
the recombination wave reaches the luminosity shell before the end of the plateau depends on the progenitor properties (e.g., it is more likely to take place in less extended and more massive progenitors, and in progenitors with a steep density profile).

Another interesting result of our analysis is that the observed plateau is not a generic property of the  propagation of a recombination wave in an expanding ionized gas. The generic property is the much slower evolution of the observed temperature once recombination starts (due to the strong feedback of the temperature drop on the thermal coupling). This fixes the observed temperature around the $R$ and $I$ bands. However, the slow decline of the bolometric luminosity is a result of the typical hydrodynamical structure of exploding stellar envelopes. Different structures can result in much faster decline rates or alternatively a rise in the bolometric luminosity.

We also find that differences in the envelope density structure contribute to the variety in the light curve decline rates observed in type-II SNe, such that progenitors with steeper density profiles generally produce SNe with more rapidly declining light curves. However, the light curve is flattened when the density profile is steep enough such that recombination reaches the luminosity shell before the entire envelope is exposed, and causes the luminosity to increase. The distribution in decline rates may account for some of the diversity seen in the light curves of type-II SNe.

Finally, the derivation of the fully analytic solution brought in this paper required the following approximations -- 
(1) assuming a power-law profile for the density and internal energy at late times; 
(2) neglecting the density dependence on the ionization fraction;
(3) approximating the ionization fraction as  a power law of the temperature;
(4) neglecting energy deposition by radioactive decay;
(5) assuming free-free absorption as the most efficient process for photon production;
(6) assuming a perfectly spherical ejecta.
However, the basic ideas of our model- separation of the envelope into successive shells and finding the recombination shell by the properties of its boundaries, can be used for more general assumptions than (1)--(6) above.
The simple equations introduced in Eq \eqref{eq:general} can be extended even further by changing the function $\eta$ such that it includes other absorption processes (such as bound-free) or by adding energy from radioactive decay. Hence, our model serves as a simple tool to examine the effect of some basic mechanisms on the evolution of SN light curves.

\section{Acknowledgments}
This research was partially supported by an ISF grant and an iCore center. We thank Roni Waldman and Eli Livne for providing us with an example of a numerical stellar envelope.


												

\appendix 

\section{The structure of the recombination shell} \label{sec:delta_acc}
The structure of the recombinaion shell can be found by solving the following diffusion equation for the energy density $\varepsilon$:
\begin{equation*}
\frac{\partial{\varepsilon}}{\partial{t}} + v \frac{\partial{\varepsilon}}{\partial x}=\frac{\partial}{\partial x}\Big(\mathcal{D}_0 \frac{\partial \varepsilon}{\partial x} \Big) ~,
\end{equation*}
where $\mathcal{D}_0$ is the diffusion coefficient.
We can assume that the energy profile in the shell is in steady state such that $\partial{\varepsilon}/\partial{t} = 0$.
When $T_{\gas}>T_{\rec}$ the diffusion coefficient is constant and can be written as $\mathcal{D}_0 = c/3 \kappa_\T \rho$. Therefore, the solution to the above equation is
\begin{equation*}
\varepsilon(T_\gas>T_\rec) = \frac{F_0}{v}\Big(1- e^{xv/\mathcal{D}_0}\Big)  + \varepsilon_\rec ~,
\end{equation*}
with $x=0$ corresponding to $T_\gas = T_\rec$, and $F_0 = L/4\pi r_\rec^2$ is the radiative flux where the diffusive flux dominates and $L$ is constant. The scale over which $\varepsilon$ changes is defined as:
\begin{equation*}
\delta(T_\gas>T_\rec) = \frac{\varepsilon}{\partial \varepsilon/ \partial x} = \frac{\mathcal{D}_0}{F_0} \varepsilon e^{-x v/ \mathcal{D}_0} ~.
\end{equation*}
In obtaining the above expression for $\delta$, we did not assume that $L$ is constant. Therefore, it is also correct for $t>t_\L$ in regions satisfying $t>t_{\text{d}}$, and is more general than the expression in Eq \eqref{eq:delta}.
We note that for $x \ll \mathcal{D}_0/v$ (which implies that $\tau \ll c/v$ and constant $L$):
\begin{equation*}
\varepsilon(T_\gas>T_\rec)\approx \frac{F_0 x}{\mathcal{D}_0} ~,
\end{equation*}
and
\begin{equation*}
\delta(T_\gas>T_\rec) = \frac{\mathcal{D}_0 \varepsilon}{F_0}~.
\end{equation*}
In terms of $T_\gas$, the expression for $\delta$ becomes:
\begin{equation*}
\delta(T_\gas>T_\rec) =  \Big(\frac{T_\rec}{T_\rsout}\Big)^{4-\beta} \Big(\frac{T_\gas}{T_\rec}\Big)^{4} \times r_\rec ~,
\end{equation*}
equivalent to Eq \eqref{eq:delta}.

For $T<T_\rec$, the diffusion coefficient is a function of $\varepsilon$, and the diffusion equation in steady state becomes:

\begin{equation*}
v \frac{\partial{\varepsilon}}{\partial x}=\frac{\partial}{\partial x}\Big[\mathcal{D}_0 \Big(\frac{\varepsilon}{\varepsilon_\rec}\Big)^{-\beta/4} \frac{\partial \varepsilon}{\partial x} \Big] ~.
\end{equation*}
Assuming that in regions where $T_\gas<T_\rec$ the diffusion term dominates over the advection term, the luminosity is constant and the solution for the above equation is
\begin{equation*}
\varepsilon(T_\gas<T_\rec) = \varepsilon_\rec \Big(1 - \frac{4-\beta}{4 ~\varepsilon_\rec}\frac{F_0}{\mathcal{D}_0}x\Big)^{(4-\beta)/4} ~.
\end{equation*}
The scale over which $\varepsilon$ changes is 
\begin{equation*}
\delta(T_\gas<T_\rec) = \frac{\mathcal{D}_0 ~\varepsilon}{F_0} \varepsilon_\rec \Big(\frac{\varepsilon}{\varepsilon_\rec}\Big)^{(4-\beta)/4}~, 
\end{equation*}
and in terms of $T_\gas$:
\begin{equation*}
\delta(T_\gas>T_\rec) = \Big(\frac{T_\gas}{T_\rsout}\Big)^{4-\beta} \times r_\rec ~,
\end{equation*}
reproducing the expression in Eq \eqref{eq:delta}.
\vspace{1cm}
\bibliographystyle{apj}	
\bibliography{recombination_revised}
\end{document}